\documentclass[a4paper,12pt]{article}
\pdfoutput=1
\usepackage{amsmath,amssymb}
\usepackage{jheppub}
\usepackage{hyperref}
\usepackage{braket}
\usepackage{graphicx}
\usepackage{mathrsfs}
\usepackage{float}
\usepackage{ulem}
\usepackage{color}

\newcommand{\be}{\begin{equation}}
\newcommand{\ee}{\end{equation}}
\newcommand{\beq}{\begin{equation}}
\newcommand{\eeq}{\end{equation}}

\newcommand{\bea}{\begin{equation}\begin{aligned}}
\newcommand{\eea}{\end{aligned}\end{equation}}
\newcommand{\ba}{\begin{align}}
\newcommand{\ea}{\end{align}}

\title{Complexity and the Bulk Volume, A New York Time Story}

                                           \author[a]{Alexandre Belin,}
                                           \author[b]{Aitor Lewkowycz}
                                           \author[c,d]{and G\'abor S\'arosi}
                                           \affiliation[a]{Institute for Theoretical Physics, University of Amsterdam \\
                                           Science Park 904, 1098XH Amsterdam, The Netherlands,}
                                           \affiliation[b]{{Stanford Institute for Theoretical Physics, Department of Physics,\\
Stanford University, Stanford, CA 94305, U.S.A.}}
                                           \affiliation[c]{David Rittenhouse Laboratory, University of Pennsylvania,\\
                                           Philadelphia, PA 19104, USA}
                                           \affiliation[d]{Theoretische Natuurkunde, Vrije Universiteit Brussels and \\ International Solvay Institutes,\\
Pleinlaan 2, Brussels, B-1050, Belgium}
                                           \emailAdd{a.m.f.belin@uva.nl}
                                           \emailAdd{lewkow@stanford.edu}
                                           \emailAdd{sarosi@sas.upenn.edu}
                                           
\abstract{
We study the boundary description of the volume of maximal Cauchy slices using the recently derived equivalence between bulk and boundary symplectic forms. The volume of constant mean curvature slices is known to be canonically conjugate to ``York time". We use this to construct the boundary deformation that is conjugate to the volume in a handful of examples, such as empty AdS, a backreacting scalar condensate, or the thermofield double at infinite time. 
We propose a possible natural boundary interpretation for this deformation and use it to motivate a concrete version of the complexity=volume conjecture, where the boundary complexity is defined as the energy of geodesics in the K\"ahler geometry of half sided sources. We check this conjecture for Ba\~nados geometries and a mini-superspace version of the thermofield double state.
Finally, we show that the precise dual of the quantum information metric for marginal scalars is given by a particularly simple symplectic flux, instead of the volume as previously conjectured.
}

\begin{document}
\maketitle

\section{Introduction}

In the recent years, quantum information theory has helped improving our understanding of quantum gravity and holography. Starting from the work of
Ryu and Takayanagi \cite{Ryu:2006bv}, where a precise holographic dual to the entanglement entropy was proposed, we now have a quite elaborate picture of how subregions in holographic theories work. In particular, we understand the duals of quantities like R\'enyi entropies, relative entropies or modular flows \cite{Dong:2016fnf,Jafferis:2015del} and the mapping between low energy operators in the bulk and in the boundary \cite{Almheiri:2014lwa}. These results were also useful for building toy models of holography using tensor networks \cite{Pastawski:2015qua,Hayden:2016cfa}. 
It is however a little mysterious why there has been so much progress in understanding simple gravitational quantities associated with boundary subregions (like codimension-2 RT surfaces) in a particular state, but 
not much has been precisely understood about duals of geometric quantities localized to codimension-1 surfaces, which are naturally associated with pure states in the boundary. While the former depend on both the state and the subregion, the latter only depend on the state so they should be somehow simpler. 

Without doubt, the most interesting quantity in this regard is the volume of the extremal surface which asymptotes to the respective boundary
Cauchy slice. In \cite{Stanford:2014jda,Susskind:2014rva}, it has been proposed that this volume measures the complexity of the quantum state of the boundary (see also \cite{MIyaji:2015mia} where it is argued that it represents fidelity susceptibility), but up to now there has not been a precise boundary description of this volume. In fact, it is still not clear how to define complexity in a generic QFT (although see \cite{Caputa:2017urj,Caputa:2017yrh,Czech:2017ryf,Jefferson:2017sdb,Chapman:2017rqy,Bhattacharyya:2018wym,Takayanagi:2018pml,Magan:2018nmu,Caputa:2018kdj,Chapman:2018hou,Ali:2018fcz} for recent progress).

The goal of this work is to understand properties of pure states in holographic CFTs, which are prepared by turning on sources in the Euclidean path integral. We are going to denote such a state generally as $|\lambda \rangle$, with $\lambda(x)$ representing a general source for our operators on half of the Euclidean manifold, over which the path integral is performed. In particular, these include states where we only change the Euclidean boundary metric. These states are characterized by a set of non-vanishing one-point functions induced by the sources which map to non-trivial backgrounds in the bulk. They are therefore the boundary duals of classical geometries. 

Our starting point will be the results of \cite{Belin:2018fxe}, where the bulk symplectic form was related with the quantum overlap between nearby states in the boundary. In terms of normalized states $|\Psi_\lambda \rangle= \frac{|\lambda\rangle}{\langle \lambda|\lambda\rangle^{1/2}}$, this relation reads as 
\begin{equation}
\label{eq:symplBerry}
\Omega(\delta_1 \lambda,\delta_2 \lambda)=\int_{\Sigma} \delta_1 \phi \delta_2 \pi - \delta_2 \phi \delta_1 \pi= i  \langle \delta_1 \Psi_\lambda| \delta_2\Psi_\lambda \rangle - (1 \leftrightarrow 2)
\end{equation}
Here, we are denoting linearized deformations by $ |\delta\Psi_\lambda \rangle \approx |\Psi_{\lambda+\delta \lambda} \rangle -|\Psi_\lambda \rangle$,  
the bulk field $\phi$ is dual to the operator sourced by $\lambda$, its conjugate momentum is $\pi$, and $\Sigma$ is a bulk initial value surface. This is the main result of \cite{Belin:2018fxe} and it gives a boundary information theoretical interpretation of the bulk symplectic form, as the Berry curvature of the parametrized family of states $|\Psi_\lambda \rangle$. 

This symplectic form can be used to express the change in the extremal volume. There is a particular deformation of the boundary sources, $\delta_Y$, which we will call the ``new York" deformation, that satisfies
\begin{equation}
\label{eq:introvolumesympl}
\Omega(\delta_Y \lambda, \delta \lambda)= \delta V_{\rm ext}.
\end{equation}
We will see that it is easy to write $\delta_Y$ in the bulk in terms of the bulk metric, however, its physical interpretation is not immediately clear. Furthermore, obtaining the respective boundary deformation is complicated and state dependent. The main goal of this paper is to understand what the boundary interpretation of this deformation is.

We will begin by exploring the physical meaning of $\delta_Y$ from the bulk point of view, by studying it in the context of ADM and York \cite{Arnowitt:1959ah,York:1972sj}. In \cite{York:1972sj}, York proposes that by foliating a geometry with constant scalar extrinsic curvature slices (York time), one can characterize the gauge invariant gravitational phase space. In the context of AdS, this gives a background dependent foliation of the Wheeler-De Witt patch (see left of Fig. \ref{fig:2}), which seems to be particularly well suited for the Hamiltonian formalism, and the respective York Hamiltonian is the volume. The new York transformation $\delta_Y$ corresponds
to doing ``half" of a translation in York time, that is either staying on the same surface but evolving the initial data or evolving the surface but keeping the initial data fixed. Doing both would be a plain diffeomorphism that is naively trivial at the boundary, but evolving only
half of the degrees of freedom acts physically on the Hilbert space. We will see that this new York transformation is conjugate to the volume with respect to an ``unconstrained" symplectic form, which only involves the physical phase space coordinates.

 We then proceed by exploring the dual of the new York transformation for three particular examples: the vacuum state, the thermofield double at infinite time, and a scalar condensate that is perturbatively close to the vacuum. For the vacuum state, it is easy to understand the new York deformation, but because
of the symmetries, it turns out that we effectively have  $\delta V=0$ for any deformation of the CFT background metric. In the boundary, the new York deformation can be written as
\begin{equation}
|0\rangle \rightarrow \delta_Y \gamma_{a b}(t_E)=i \text{sign}(t_E) \delta_{a b}
\end{equation}
where $\gamma_{ab}$ is the boundary background metric and $t_E$ is Euclidean time.
  
  Surprisingly, the late time TFD state is the next simplest example. In this situation, the boundary deformation is better understood in terms of a Weyl rescaling plus a change of coordinates:
\begin{equation}
|TFD,t=\infty\rangle \rightarrow (\delta_Y t,\delta_Y x^j ,\delta_Y \Phi)=\left(\frac{d-1}{d} t,-\frac{1}{d} x^j, -\frac{i}{2}\right) 
\end{equation}
where $\Phi$ is the conformal factor of the boundary metric. Using this deformation, we can reproduce the well known late time growth of the volume in the TFD state \cite{Susskind:2014rva,Stanford:2014jda} from the symplectic form. The previous equation also implicitly has a $\text{sign} t_E$, but it is valid for all Lorentzian times as we will explain. 

As a third example, we will consider a state obtained by turning on a scalar operator with a perturbatively small source. This corresponds to a scalar condensate in the bulk, and we will consider the leading order backreaction of this condensate, which can lead to a finite change in the volume. We will show that the boundary new York transformation associated to this state involves exclusively the source for the scalar operator, and not the background metric. 

We will also explain the relation between our story and the fidelity susceptibility of \cite{MIyaji:2015mia} (see also \cite{Bak:2015jxd,Trivella:2016brw,Banerjee:2017qti} for some recent discussions about this quantity). In our context, the fidelity susceptibility can be understood as the symplectic pairing between a constant source deformation $\delta_c \lambda=\lambda$ and a sign deformation $\delta_s \lambda = i \text{sign}(t_E) \lambda$.  For marginal operators, it is easy to show that
\begin{equation}
G_{\lambda \lambda} \propto \Omega(\delta_s \lambda, \delta_c \lambda)=\lambda \int_{\Sigma} \delta_s \pi,
\end{equation}
that is, the dual of the fidelity susceptibility is the integral of the sign deformed momenta on a bulk Cauchy slice. While this is generally different from the volume, we will show that for the case of the vacuum and the thermofield double at late times, it indeed agrees with the volume, simply because the respective sign deformed momenta is constant over the maximal slice.

In the last part of the paper, we attempt to connect this story to the complexity=volume conjecture of \cite{Susskind:2014rva,Stanford:2014jda}. In our setup, we have a natural notion of distance in the space of Euclidean sources. This distance is essentially coming from the pull-back of the Fubini-Study metric to the space of sources, using the Euclidean path integral states. It seems sensible to define a notion of complexity between two path integral states from the kinetic energy in this geometry, that is
\begin{equation}
{\cal C}=\int_{s_i}^{s_f} ds g_{a b}(\lambda) \dot{\lambda}^a(s) \dot{\lambda}^a(s) , ~~~~  \lambda(s_{i,f})=\lambda_{i,f}
\end{equation}
where $\lambda_{i,f}$ are sources corresponding to the initial and final states, and $g_{a b}$ is schematically given by the connected two point function $\partial_{\lambda^a} \partial_{\lambda^b} \log \langle \lambda | \lambda \rangle$.
We conjecture that ${\cal C}$ computes the extremal volume. One motivation for this conjecture is that this would give a natural boundary interpretation for the new York transformation $\delta_Y$ in \eqref{eq:introvolumesympl}, as being related to the tangent vector of this geodesic at the end point, that is $\dot \lambda(s_f)$. We check this proposal for two examples: Ba\~nados geometries that are close to the vacuum, and a mini-superspace version of the time dependent TFD state. In the former example, this definition coincides precisely with the volume computed by \cite{Flory:2018akz}, and in the latter example, we find qualitative agreement with the expected
behaviour for the volume in holographic theories. This also gives an example of calculating a complexity-like quantity in field theory without relying on weak coupling.

The structure of the paper is the following. In section \ref{sec:review}, we review and expand the content of \cite{Belin:2018fxe}, setting up the notation for the rest of the paper. In section \ref{sec:adm}, we review York time and explore the bulk interpretation of the new York deformation. Section \ref{sec:vacuumdeformation} and \ref{sec:inftfd} explore explicitly the concrete examples of the vacuum state, the scalar condensate and the thermofield double state at late times. In section \ref{sec:complexity}, we contextualize our findings in light of the complexity=volume conjecture and we finally close with section \ref{sec:disc} with general comments and future directions.

\section{Equality of bulk and boundary symplectic forms}
\label{sec:review}
 
 \subsection{Review and notation}
 
We start by reviewing the result of \cite{Belin:2018fxe}. Given a set of coherent states parametrized by some phase space, there is a canonical way of recovering the symplectic stucture.   This can be thought of as running ``backwards" the usual quantization procedure. There, one starts with a phase space, which is a symplectic manifold. Quantization then requires a choice of dividing the phase space into coordinates and momenta, since a quantum wave function can only depend on half of the phase space coordinates. In a more mathematical language, this is equivalent with choosing an almost\footnote{We will be concerned with local aspects of phase space and will not discuss whether these structures are globally well defined.} complex structure that is compatible with the symplectic form. This gives phase space the structure of a K\"ahler manifold.

Now suppose instead that we are given a quantum Hilbert space with a set of ``candidate" coherent states, parametrized by some complex coordinates $(\lambda,\lambda^*)$. Can we recover this K\"ahler stucture from the inner product of this Hilbert space? The answer is yes.\footnote{Some properties of this ``dequantization" were studied before in \cite{rawnsley1977coherent} in a different context. The role of the K\"ahler geometry of the projective Hilbert space in quantum mechanics was also studied before, see e.g. \cite{Ashtekar:1997ud} and references therein.} First, we fix the complex structure by requiring that conjugation on $\lambda$ means the same as on the Hilbert space. One way of achieving this is to just ask for holomorphic embeddings, that satisfy
\bea
\label{eq:holostates}
\partial_{\lambda^*}|\lambda\rangle=0, && \quad && \partial_\lambda \langle \lambda|=0.
\eea
Such states are neccessarily unnormalized. Luckily, the inner product gives rise to a canonical metric on the projective Hilbert space, the Fubini-Study metric
\be
ds^2= \frac{\braket{\delta\psi|\delta\psi}}{\braket{\psi|\psi}}-\frac{|\braket{\delta\psi|\psi}|^2}{\braket{\psi|\psi}^2} \,.
\ee
Pulling this metric back with a map that satisfies \eqref{eq:holostates}, one obtains the line element
\beq
ds^2 = \partial_\lambda \partial_{\lambda^*}\log \langle \lambda | \lambda \rangle d\lambda d\lambda^*,
\eeq
which defines a K\"ahler potential $\mathcal{K}=\log \langle \lambda | \lambda \rangle$. The symplectic form is then just the K\"ahler form of this potential
\beq
\Omega = i \partial_\lambda \partial_{\lambda^*}\log \langle \lambda | \lambda \rangle d\lambda \wedge d\lambda^* .
\eeq
It is worth noting that this object is also the Berry-curvature two form associated with the parameter space $(\lambda,\lambda^*)$. Indeed, this expression can be written in terms of the normalized family of states $|\Psi_\lambda \rangle = (\langle \lambda | \lambda \rangle)^{-1/2}|\lambda \rangle$ as $\Omega = i d\langle \Psi_\lambda |d|\Psi_\lambda \rangle$ which is the same as the r.h.s. of \eqref{eq:symplBerry} when evaluated on explicit variations.

\begin{figure}[h!]
\centering
\includegraphics[width=0.65\textwidth]{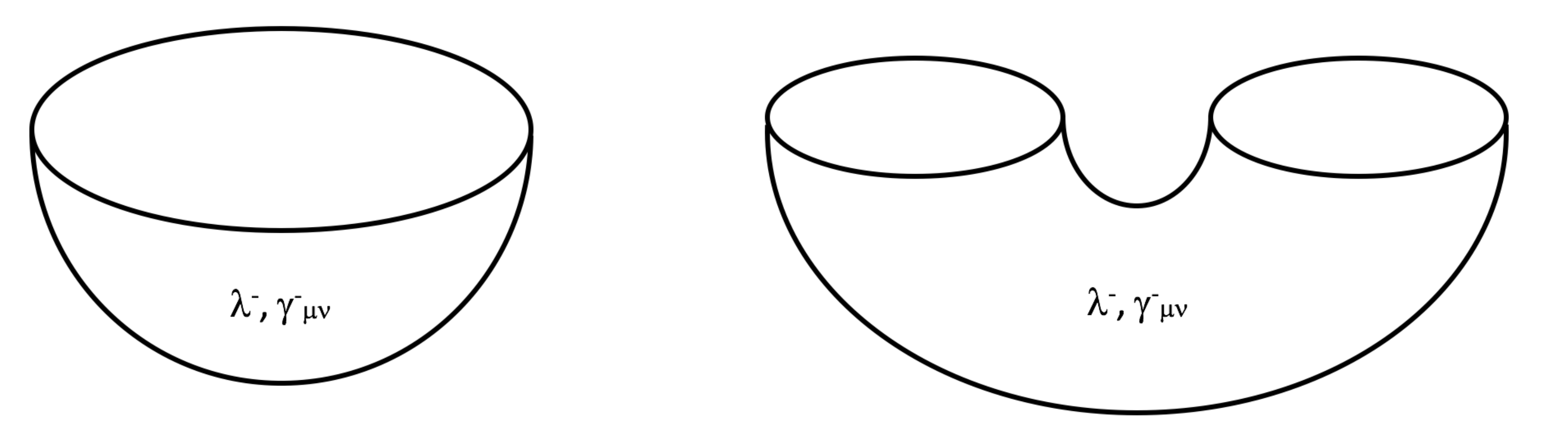}
\caption{Two euclidean path integral states with sources prepared on manifolds of different topology (the hemisphere and the cylinder). On the left, the state lives in a single copy of the CFT Hilbert space $\mathcal{H}$ while on the right, it lives in $\mathcal{H}\times\mathcal{H}$.}
\label{fig:0}
\end{figure}

In quantum field theory, a natural set of states is obtained by path integrating over half of the Euclidean manifold, while turning on sources for certain operators. Formally, we can write such states with the Euclidean time ordering symbol $T$ as
\bea
\label{eq:pathintstates}
|\lambda\rangle &= Te^{-\int_{t_E<0} dt_E d^{d-1}\vec{x} \lambda^-(t_E,\vec{x}) O(t_E,\vec{x})}|0\rangle , && \langle \lambda | &= \langle 0|Te^{-\int_{t_E>0} dt_E d^{d-1}\vec{x} \lambda^+(-t_E,\vec{x}) O^\dagger (t_E,\vec{x})}.
\eea
The functions $\lambda^\mp$ are defined for the $t_E<0$ half of the Euclidean manifold, and they are complex. We think of $\lambda^-(x)$ as the holomorphic coordinate, and its complex conjugate $\lambda^+(x)\equiv[\lambda^-(x)]^*$ as the antiholomorphic coordinate on this space. The states are such that ``ket" states only depend on holomorphic, while the ``bra" states only depend on antiholomorphic sources, similarly as in \eqref{eq:holostates}. Whenever we write $\lambda$ without a superscript, we refer jointly to the sources $\lambda^\mp$. With a slight abuse of notation, we will also use $\lambda$ for the joined source profile
\bea
 \lambda(x) &=
\begin{cases} \lambda^-(t_E^x, \vec{x}) & t^x_E<0\\
\lambda^+(-t_E^x,\vec{x}) & t^x_E>0,
\end{cases}
\label{eq:lambdatilde}
\eea
which is now defined over the entire Euclidean manifold.
Note that when the sources are real and independent of $t_E$, we can think of the states \eqref{eq:pathintstates} as the ground states of a deformed Hamiltonian
\beq
H=H_{0}+\int d^{d-1}\vec{x} \lambda(\vec{x})O(0,\vec{x}).
\eeq
Note the background on which the $\ket{\lambda}$ states are defined may be the hemi-sphere or something more involved like the cylinder with two boundaries which prepares the thermofield-double state (see Fig. \ref{fig:0}). 

Before moving on, let us also briefly lay out some of the notation that we will use for functionals. We reserve $\delta$ for describing tangent vectors in the space of field configurations, $\delta_f=\int dx \delta f(x) \frac{\delta}{\delta f(x)}$. We will never write formal differentials dual to functional derivatives, rather we describe differential forms on field space by their action on tangent vectors, as is usually done in discussions involving covariant phase space methods. The symbols $\delta^-$, $\delta^+$ will refer to holomorphic and antiholomorphic variations that are only defined on half of Euclidean space, and we use $\delta$ either to refer to these jointly or for the variation of the joined profile \eqref{eq:lambdatilde}, in which case $\delta= \delta^-+\delta^+$.

The K\"ahler potential associated to the family of states \eqref{eq:pathintstates} is obtained by performing the Euclidean path integral that calculates the norm $\langle \lambda | \lambda \rangle$. It is just given by the generator of connected Feynman diagrams
\bea
\mathcal{K} &= \log \langle \lambda | \lambda \rangle =\log Z[ \lambda], && \quad &&
\label{eq:kahlerpotqft}
\eea
Note that the source $\lambda$ in this generating function is not arbitrary, it has to be invariant under Euclidean time reflection combined with a complex conjugation (or $Z_2+C$). 
The K\"ahler metric and the K\"ahler form are obtained from the second variations of this object and are therefore controlled by the connected two point function $G^c_{ \lambda}(x,y)=\langle O(x) O(y) \rangle_{ \lambda}^{conn}$ in the presence of the source $\lambda$. For example, the K\"ahler form evaluated on field variations $\delta_j  \lambda \equiv (\delta_j \lambda^-, \delta_j \lambda^+)$ reads as
\bea
\label{eq:qftkahler}
\Omega (\delta_1  \lambda,\delta_2  \lambda)  &=i(\delta_1^+ \delta_2^--\delta_1^- \delta_2^+)\log Z[\lambda] \\
&=i \int_{\substack{t^x_E>0\\t^y_E<0}} dx   dy G^c_{ \lambda}(x,y) [\delta \lambda_1^+(x^T) \delta \lambda_2^-(y)-\delta \lambda_1^-(y) \delta \lambda_2^+(x^T)],
\eea
 where we are using the shorthand $x^T=(-t_E^x, \vec{x})$ for reflection in Euclidean time. Note that the two legs of the two point function are integrated on opposite half sides of the Euclidean manifold. The K\"ahler metric has a similar expression, but it contains the symmetrized combination of the variations, and there is no $i$ in front. The relation between the K\"ahler metric and the K\"ahler form is, as usual, given by the complex structure, $g(\delta  \lambda, \delta \lambda)=\Omega(\delta  \lambda, J[\delta  \lambda])$, where the complex structure acts on the sources as
 \bea
 \label{eq:J}
 J[\lambda^-(x)]=i \lambda^-(x), && \quad && J[\lambda^+(x)]=-i\lambda^+(x).
 \eea
 Finally, we can also write expression \eqref{eq:qftkahler} in a more local form
 \beq
\label{eq:canonicalpair}
\Omega (\delta_1 \lambda,\delta_2  \lambda) = i \int_{t_E>0} dx(\delta \lambda_1^+\delta^-_2 \langle O \rangle-\delta \lambda_2^+\delta_1^- \langle O \rangle).
\eeq
This expression also has a more intuitive interpretation, where one views the K\"ahler form as a symplectic form, pairing the source and expectation value of the operators. One can therefore think of the sources and VEVs as canonically conjugate pairs. 

Before moving on, it should be noted that in complete generality, the states \eqref{eq:pathintstates} cannot be defined directly in the continuum. For example, inserting sources for irrelevant operators lead to divergences in correlation functions once one expands the exponential. One should therefore view these states as formal power series,  only properly defined when working at finite UV cutoff. We defer a more in-depth discussion of this issue for the discussion section but nevertheless, there are some theories for which the situation is better. One such example is large $N$ CFTs where we only source single trace operators (it is natural to think about single trace operators as the ones making up the classical phase space in the large $N$ limit \cite{Yaffe:1981vf}).

In the context of holography, turning on boundary sources is equivalent to changing the boundary conditions for the bulk fields. In this way, using the standard dictionary, there is a natural correspondence between semi-classical geometries and  $\ket{\lambda}$ states.
These two theories are related by the standard dictionary: 
\begin{eqnarray}
\langle \lambda|\lambda\rangle=Z_{CFT}[ \lambda]=e^{-S_{\rm grav}[ \lambda]}.
\end{eqnarray}
We therefore have that the K\"ahler potential \eqref{eq:lambdatilde} in holographic theories is given by the on-shell gravitational action $\mathcal{K}=-S_{\rm grav}[ \lambda]$. In this context, we may also take this to simply be the definition of the K\"ahler potential without worrying about how to construct explicitly the $|\lambda\rangle$ states in the field theory. 

\begin{figure}[h!]
\centering
\includegraphics[width=0.65\textwidth]{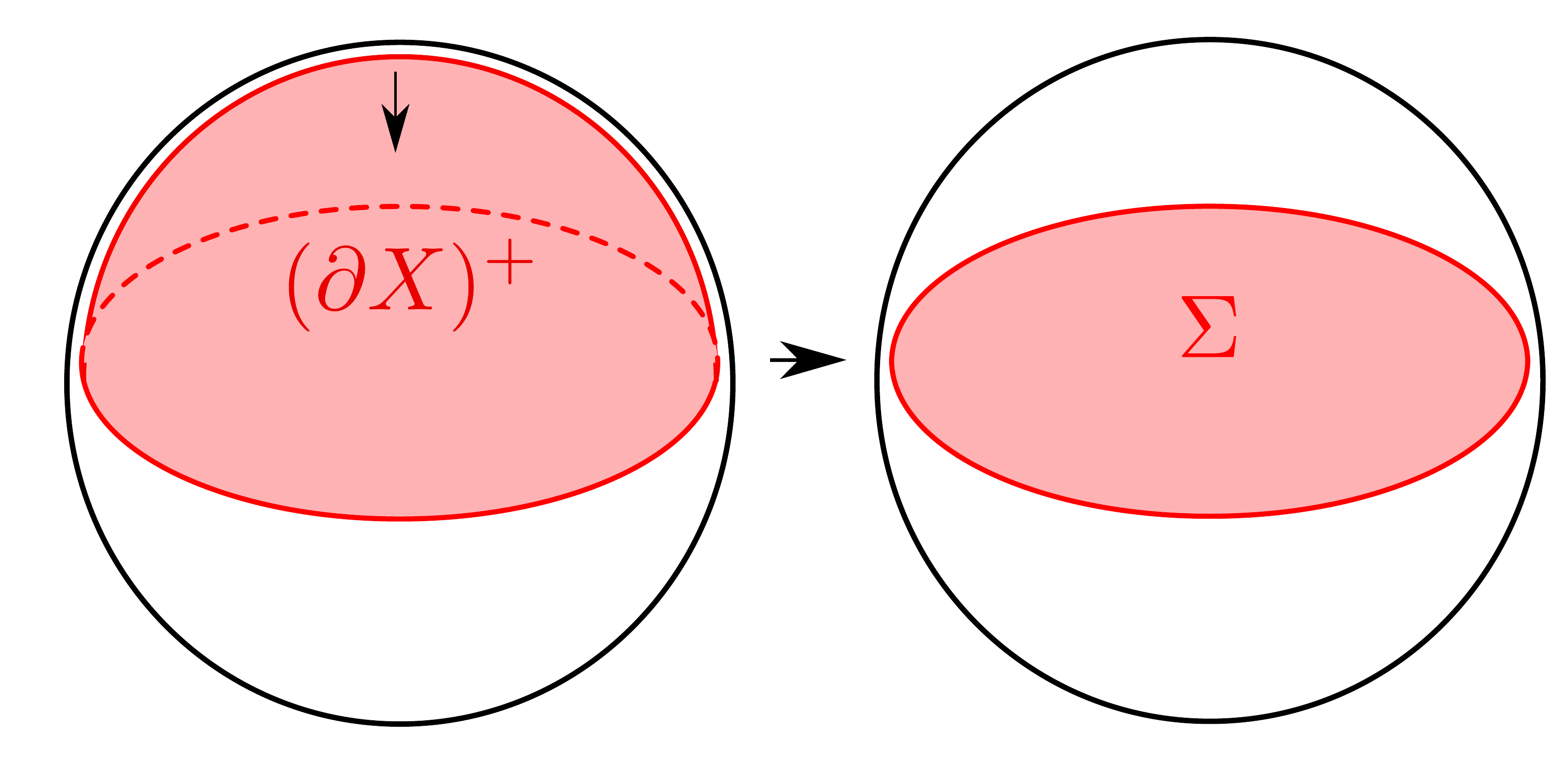}
\caption{Using conservation we can push the symplectic form from half of the Euclidean boundary to an arbitrary bulk slice $\Sigma$ anchored to the boundary at $t_E=0$.}
\label{fig:1}
\end{figure}

This leads to the main observation in \cite{Belin:2018fxe}, namely that for holographic field theories, the standard dictionary relates \eqref{eq:canonicalpair} with the bulk symplectic flux along half of the Euclidean boundary. The bulk symplectic flux is defined via Wald's procedure \cite{Crnkovic:1987tz,Lee:1990nz}, i.e. by first defining the presymplectic form $\theta$ from the boundary term in the variation of the bulk Lagrangian
\beq
\delta \mathcal{L}_{\rm bulk}d^{d+1}x =-E_\phi \delta \phi d^{d+1} x + d \theta(\phi,\delta \phi), 
\eeq
where $E_\phi$ is are the bulk equations of motion, and then defining the $d$-form
\beq
\omega_{\rm bulk}(\phi,\delta_1 \phi, \delta_2 \phi)= \delta_1 \theta(\phi,\delta_2 \phi)-\delta_2 \theta(\phi,\delta_1 \phi),
\eeq
whose integral on a codimension one surface gives the symplectic flux.
Since this symplectic flux is conserved on shell, we can push the surface into the Euclidean bulk. See Fig. \ref{fig:1} for an illustration of this. Note that for the bulk symplectic form to be equal to the boundary one, we only need for the union of the half-boundary and the bulk Cauchy slice to be a closed manifold, but for the rest of the paper we will focus on the case where the topology of the  bulk Cauchy slice and the half boundary are the same.\footnote{Some examples of when this is not the case are the TFD state below the Hawking-Page transition or the AdS soliton \cite{Horowitz:1998ha,Belin:2018jtf}.}

During this procedure, one needs to solve for the bulk geometry that has boundary conditions $\lambda$ for the bulk fields. On slices that continue nicely to Lorentzian signature, this gives a prescription for associating bulk initial data to the state $|\lambda \rangle$, as originally proposed in \cite{Skenderis:2008dg} and further studied in \cite{Marolf:2017kvq}. This Lorentzian initial data is real precisely if $\lambda^+$ is the conjugate of $\lambda^-$. This continuation gets rid of the $i$ in \eqref{eq:canonicalpair} and we obtain an equality between the Berry curvature on the space of half sided sources and the bulk symplectic form for the corresponding Lorentzian initial data on a Cauchy slice $\Sigma$
\bea
\label{eq:mainresult}
\Omega (\delta   \lambda_1,\delta \lambda_2) =  \int_{\Sigma} \omega_{\rm Lor}(\phi,\delta \phi_1,\delta \phi_2).
\eea
Similarly, the complex structure \eqref{eq:J} provides a quantum polarization of the bulk phase space, which corresponds to separation of on-shell field variations into positive and negative energy modes, whenever a time like Killing vector is available. The K\"ahler metric corresponds to the Klein-Gordon product in the bulk \cite{Belin:2018fxe}. Note that while symplectic forms in the space of sources have appeared before in holography (see e.g. \cite{Shyam:2017qlr} and references therein), these are different from our construction, since they correspond to bulk symplectic forms on time-like slices, and they vanish on-shell.

The conservation of the bulk symplectic flux can also be rephrased as the formula for the connected boundary two point function
\beq
G^c_{\lambda}(x,y) = i \int_{\Sigma} \omega_{\rm bulk}\big(K_E(Y|x),K_E(Y|y)\big),
\eeq
where $K_E(Y|x)$ is the Euclidean boundary to bulk propagator in the background defined by the boundary condition $\lambda$.
A special case of this formula has played a role in obtaining nonlinear gravitational EOMs from equating bulk and boundary relative entropies \cite{Faulkner:2017tkh,Haehl:2017sot}.

\subsection{Example: conserved charges}

A simple check of the relation \eqref{eq:mainresult} can be performed by recovering Wald's conserved charges. Consider for example the state $|\lambda,t \rangle = e^{-i t H}|\lambda\rangle$, where we evolve a little in Lorentzian time, but we think about this as turning on the Hamiltonian with a purely imaginary source. The boundary symplectic form \eqref{eq:qftkahler} between this imaginary source, and an arbitrary variation is then given by
\beq
\Omega(\delta  \lambda, \delta \lambda_{\delta t}) = (\delta^-+\delta^+)\frac{\langle \lambda |H|\lambda \rangle}{\langle \lambda | \lambda \rangle}\equiv  \delta \langle H \rangle.
\eeq
This shift in time can be sourced by changing the boundary metric by a diffeomorphism $\delta \gamma_{ab} = \nabla_{(a} \xi_{b)}$, if at the $t=0$ slice $\xi=\partial_t$. Similarly, in the bulk we can consider any bulk diffeomorphism $\zeta$ such that near the boundary $\zeta$ approaches $\xi$ and we have that
\beq
\label{eq:Hamiltondef}
 \delta \langle H \rangle = \int_{\Sigma} \omega_{\rm Lor}(\phi,\delta\phi,\mathcal{L}_\zeta \phi).
\eeq
This is the usual covariant phase space definition of conserved charges and thus provides a simple explicit check of the equality between symplectic forms.  
Of course, this same discussion applies to any charge, $\Omega(\delta \lambda,\delta \lambda_{\delta \mu})=\delta \langle Q\rangle$, with $\mu$ the conjugate to the charge $Q$.

\subsection{Example: modular flow}

Another interesting deformation of the state one can do is modular flow. Consider the following state
\beq
|\lambda,s\rangle = e^{-i s ( K_{0}-K_\lambda)}|\lambda\rangle 
\eeq
where $K_{0}=-\log \rho_{0}$ and $K_{\lambda}=-\log \rho_{\lambda}$ are the modular Hamiltonians of the vacuum and the state $|\Psi_\lambda \rangle$ respectively, associated to some spatial subregion. This deformation might be interpreted as turning on a source for the metric in terms of the replica trick, where we insert a (possibly complex) conical deficit around the entangling surface. 

We can calculate the boundary symplectic form at $s=0$, and it is basically the same as in \eqref{eq:Hamiltondef} with the relative modular Hamiltonian instead of the Hamiltonian:
\bea
\Omega(\delta  \lambda,\delta \lambda_{\delta s}) & = (\delta^-+\delta^+)\frac{\langle \lambda|K_{0}-K_\lambda|\lambda \rangle}{\langle \lambda |\lambda \rangle}& = \delta S(\rho_\lambda || \rho_{0}),
\eea
where $S(\rho_\lambda || \rho_{0})$ is the relative entropy and, by definition, the $\lambda$ variation does not act on the $K$'s. Comparing with the result \eqref{eq:mainresult} implies that we should have
\beq
\delta S(\rho_\lambda || \rho_{0}) = \int_\Sigma \omega_{\rm bulk}(\phi,\delta \phi,\partial_s \phi),
\eeq
where $\partial_{s}\phi$ is the variation of the bulk field under modular flow.

In general, this can be very complicated, but is computable in principle using free field techniques in the bulk \cite{Faulkner:2017vdd}. However, close to the vacuum state, we can compute it explicitly. More concretely, consider a classical state $|\Psi_\lambda\rangle$ which can be treated in terms of bulk effective field theory on top of the vacuum, in which case it is expected to correspond to a bulk coherent state \cite{Botta-Cantcheff:2015sav}
\beq
|\Psi_\lambda \rangle = U_\lambda |0\rangle, \quad U_\lambda = \exp \Big(i\int(\langle \phi\rangle_\lambda \hat \pi - \langle \pi \rangle_\lambda \hat \phi) \Big),
\eeq
here $\hat \pi$, $\hat \phi$ are bulk effective field theory operators for the momenta and the fields. Since we would like to avoid considering backreaction, we will restrict to the situation where the bulk stress tensor in the $\lambda$ state is small in units of $G_N$ (in the presence of gravitons, by stress tensor we mean the Einstein tensor expanded to quadratic order around the background geometry). If the field expectation value is classical, this implies that we stick to calculations to second order in $\lambda$, or in other words, we pick up the Fisher information part from the relative entropy \cite{Lashkari:2015hha}. We can take the commutator of the fields with the modular Hamiltonian by using the bulk modular Hamiltonian \cite{Jafferis:2015del}. 

Using that $U_\lambda \hat \phi U_\lambda = \hat \phi - \langle \phi \rangle_\lambda$ and $K_\lambda = U_\lambda K_{0} U_\lambda^{-1}$ we have that
\bea
\partial_{s}\phi &= \langle 0|U_\lambda^{-1} [K_{0}-K_\lambda,\hat \phi] U_\lambda |0\rangle \\
&=  \langle 0|U_\lambda^{-1} [K_{0},\hat \phi] U_\lambda |0\rangle -\langle 0| [K_{0},\hat \phi]  |0\rangle.
\eea
The action of $K_{0}$ is local if the boundary subregion is a single ball shaped region and in this case one has $[K_{0},\hat \phi] =\mathcal{L}_\xi \hat \phi$, where $\xi$ now is now the bulk vector field generating the bulk vacuum modular flow inside the entanglement wedge \cite{Casini:2011kv}. Therefore, in the special case of ball-shaped regions and bulk effective field theory coherent states one has
\bea
\delta S(\rho_\lambda || \rho_{0}) &= \int_{\Sigma_E} \omega_{\rm bulk}(\phi,\delta \phi,\mathcal{L}_\xi \Delta\phi), \\
\Delta \phi &= \langle \hat \phi \rangle_\lambda - \langle \hat \phi \rangle_{0}.
\eea
where $\Sigma_E$ is the bulk region between the boundary and the RT surface.
As explained before, if the field is a classical field, we have to restrict to second order in $\lambda$, which implies that we only keep the linear in $\lambda$ term in the field variation $\Delta \phi=\delta_{\lambda }\phi$. In this case, $\delta S(\rho_{\lambda}||\rho_0)|_{O(\lambda^2)}=S(\rho_{\lambda}||\rho_0)|_{O(\lambda^2)}$ and this way we have recovered the relation for relative entropy derived in \cite{Lashkari:2016idm} and used recently in \cite{Faulkner:2017tkh,Haehl:2017sot} to obtain nonlinear gravitational equations from entanglement data in CFTs. The point of this section was not to provide a generalization of this but instead to explain that the somewhat mysterious appearance of the symplectic form in this relation is a natural consequence of the correspondence \eqref{eq:mainresult} between the boundary and bulk symplectic forms.

\subsection{Relation to fidelity susceptibility}

It was proposed in \cite{MIyaji:2015mia} that the fidelity susceptibility (or information metric) is dual to the volume of an extremal slice. For marginal deformations in a CFT, these authors derive that the information metric is given by
\beq
\label{eq:infmetric}
G_{\lambda \lambda} = \frac{1}{2} \int_{t_E^x<0}dx\int_{t_E^y>0}dy \langle O(x) O(y) \rangle,
\eeq
where $O$ is the marginal operator that we are deforming by. Notice that this can be obtained from the boundary symplectic form \eqref{eq:qftkahler} by setting $\delta_1 \lambda^\mp(x) \equiv \delta_c \lambda^\mp =1$ and $\delta_2 \lambda^\mp(x) \equiv \delta_s \lambda^\mp = \mp i$. In other words, we can think about this as the K\"ahler norm of a real, constant source deformation, since $\delta_s \lambda=J[\delta_c \lambda]$, where $J$ is defined in \eqref{eq:J}. Therefore, \eqref{eq:mainresult} tells us that in the bulk we can write it as 
\beq
\label{eq:infmetricmiddle}
G_{\lambda \lambda} = \frac{1}{2} \int_{\Sigma} d^d x (\delta_c \pi \delta_s \varphi - \delta_s \pi \delta_c \varphi ), \quad \varphi\equiv \phi|_\Sigma, \quad \pi \equiv \sqrt{|h|} \partial_n \phi|_{\Sigma},
\eeq
that is, the bulk symplectic form of the marginal scalar, evaluated on the deformations obtained by solving into the bulk the boundary condition variations 
\bea
\label{eq:infmetricdeformations}
\delta_c \lambda & =1, && \quad && \delta_s \lambda = -i \text{sign}(t_E)
\eea
using the linearized equations of motion. This sign deformation $\delta_s  \lambda$ will appear many times later in this work. The $\partial_n$ in \eqref{eq:infmetricmiddle} refers to the normal derivative to the Cauchy slice $\Sigma$.

We can further simplify this expression using that $\delta_c \lambda$ is really a constant deformation and that the marginal scalar is dual to a massless field in the bulk. Since the equation $\nabla^2 \delta_c \phi=0$ is solved by $\delta_c \phi=1$, we have that $\delta_c \varphi=1$ and $\delta_c \pi=0$, and the information metric is given by
\beq
\label{eq:infmetricbulk}
G_{\lambda \lambda} = -\frac{1}{2} \int_\Sigma d^d x\sqrt{|h|} \partial_n\delta_s \phi,
\eeq
 and $\partial_n\delta_s \phi$ is given explicitly in terms of the Euclidean boundary to bulk propagator as
\beq
\partial_n\delta_s \phi(x \in \Sigma) =  \int_{t_E^y<0} d^d y [\partial_n K_E(x|y)+\partial_n K_E(x|y)^*].
\eeq

 Note that in this expression for the information metric, the Cauchy slice $\Sigma$ is an arbitrary slice anchored at $t=0$ because the symplectic flux is conserved. We could say that it gives the volume of the slice which has $\partial_n\delta_s \phi  =const$. One can check, using results in \cite{Marolf:2017kvq}, that if the background is vacuum AdS, then on the $t=0$ slice one has
\begin{eqnarray}
\partial_n\delta_s \phi|_{\text{vac}} = 1 \quad \rightarrow \quad G_{\lambda \lambda}|_{\text{vac}}= -\frac{1}{2} V.
\end{eqnarray} 

While this is true for the extremal slice in the vacuum state, it will not be true in general for other states (or other slices). 
We will explore this symplectic flux in a little more detail for the time evolved thermofield double state in section \ref{sec:tfdinfmetric}, which will explain why the exact time dependence of the information metric, obtained in \cite{MIyaji:2015mia}, is slightly different from that of the extremal volume, but why it still captures the right linear growth at late times.

 \section{Volume from the symplectic form}
 \label{sec:adm}
 
 \subsection{Volume as a Hamiltonian}
 
Let us start by summarizing some facts about the phase space in Einstein gravity. 
As discussed before, the canonical structure can be read off from the variation of the gravitational action $S_{\rm grav}$ on a spacetime region $M$ (with the appropriate GHY boundary term added)
\beq
\delta S_{\rm grav} = \int_{\partial M} \pi^{ab}\delta h_{ab} + (\text{bulk eom}).
\eeq
Here, $h_{ab}$ is the induced metric on the codimension one surface $\partial M$, and the canonical momentum reads as
\beq
\pi^{ab} = \sqrt{|h|}(K^{ab}-h^{ab}K),
\eeq
where $K_{ab}$ is the extrinsic curvature of the surface and $K$ is its trace. The $h_{ab}$ and the $\pi^{ab}$ are the conjugate variables on an initial data surface. In terms of these variables, the gravitational symplectic form is 
\begin{eqnarray}
\Omega(\delta_1 , \delta_2 )=\int_{\Sigma} (\delta_1 \pi^{a b} \delta_2 h_{a b} -\delta_2 \pi^{a b} \delta_1 h_{a b}) \,,  \label{eq:sympgrav}
\end{eqnarray}
which coincides with the covariant symplectic form of \cite{Crnkovic:1987tz} as was shown for example in \cite{Lee:1990nz}.
 Since gravity is a gauge theory, arbitrary $( h_{a b},  \pi^{a b})$ are not necesarily good coordinates in phase space, they have to satisfy the constraints\footnote{For the reader's convenience, we summarize the equations of motion in the ADM formalism in Appendix \ref{app:adm}.}
\bea
 \nabla^a K_{ab}-\nabla_b K^a_a &=0, \\
K_{ab}K^{ab}-(K)^2-R_{d}-d(d-1) &=0,
\eea
where $R_d$ is the Ricci scalar of the induced metric $h_{ab}$.
The first one is called the momentum constraint (which gives $d$ equations), and it is associated to diffeomorphisms inside the surface. This is no different than Gauss's law in ordinary gauge theory and can be dealt with in a similar manner, e.g. by fixing a gauge. The second constraint is called the Hamiltonian constraint, and it is associated to diffeomorphisms that change the initial value surface. These have no analogue in an ordinary gauge theory, but they can be gauge fixed by a procedure due to York \cite{York:1972sj}. The idea is to separate the induced metric into a conformal metric and a scale, and use the Hamiltonian constraint to solve for the scale. 

The scale is captured by the volume element $\sqrt{|h|}$ and the conformal metric is defined as
\beq
\bar h_{ab}= |h|^{-\frac{1}{d}} h_{ab},
\eeq 
and it is by construction invariant under rescalings of $h_{ab}$. The canonical structure in these new variables can be read off by rewriting $\pi^{ab}\delta h_{ab}=\pi_V \delta \sqrt{|h|}+\bar \pi^{ab} \delta \bar h_{ab}$, where
\bea
\pi_V &= \frac{2(1-d)}{d} K, &&
\bar \pi^{ab} &= |h|^{\frac{1}{d}+\frac{1}{2}}(K^{ab}-\frac{1}{d}K h^{ab}),
\eea
that is, the conjugate of the conformal metric is the traceless part of the extrinsic curvature, while the conjugate to the volume density is the trace of the extrinsic curvature.

We can now think about the Hamiltonian constraint as a differential equation for the volume density. We can choose a constant mean curvature (CMC) slicing, defined as a slicing where $K$ is constant on each slice, in which case this is called the Lichnerowitz equation.\footnote{While it is not essential for this paper, it would be interesting to establish the existence of this slicing of the WdW patch in asymptotically AdS geometries, maybe using the ideas of \cite{witten2017,Couch:2018phr}.} It admits a unique solution both in flat space \cite{York:1972sj} and in AdS \cite{witten2017}. This solution depends on the remaining variables, therefore we can think about the volume density as a functional
$\sqrt{|h|}=\sqrt{|h|}(\bar h_{ab},\bar \pi^{ab},K)$. 

In a CMC slicing, $K$ is just a number and it parametrizes the slices, so we think about it as time, while $\bar h_{ab},\bar \pi^{ab}$ are the remaining independent phase space coordinates. The volume $V=\int \sqrt{|h|}$ in this slicing should be thought of as the Hamiltonian. One way to see this, is to note that in classical mechanics, if we vary not just the end point, but the end time in the on-shell action, we get
\beq
\label{eq:varytime}
\delta S(q,t) = \delta t \partial_t S + \delta_{\rm endpt}S = -\delta t H + p \delta q.
\eeq
Since $K$ should be thought of as a time for gravity, the analogue of $S(q,t)$ is obtained by fixing $K$ instead of $\sqrt{|h|}$ at the boundary (which was considered recently in \cite{Witten:2018lgb}). This requires a change in the GHY boundary term, which amounts to a shift by the total variation $-\delta(\pi_V \sqrt{|h|})$ in the presymplectic form, which puts the variation of the on-shell gravitational action into the same form as \eqref{eq:varytime}, with the volume being the Hamiltonian. Adding this boundary term does not change the symplectic form \eqref{eq:sympgrav}. Therefore, we may think of general relativity as a theory with a time dependent Hamiltonian, and Wald's symplectic form as living on ``extended phase space", extended by the coordinates $(t,H)$. In particular, the analogue of it in classical mechanics would be
\begin{equation}
\label{eq:extendedsympl}
\omega=\delta_1 q \delta_2 p +\delta_2 t \delta_1 H - (1 \leftrightarrow 2)
\end{equation}
evaluated with the constraint $H=H(q,p,t)$. This extended symplectic form is often considered in the discussion of time dependent Hamiltonians, see e.g. \cite{struckmeier2002canonical,struckmeier2005hamiltonian} and references therein.

\subsection{The ``new York" transformation}

In order to gain a boundary understanding of the volume, we want to express it with the use of the symplectic form. This is easy to do from the classical mechanics analogue \eqref{eq:extendedsympl}, we just need to set $\delta_1 t=1$, $\delta_1 H=\delta_1 q=\delta_1 p=0$ and leave $\delta_2$ arbitrary to get $\delta_2 H$. However, since $H$ is ultimately a function of $q,p$ and $t$, we can only set $\delta H=0$ when $\partial_t H=0$, i.e. where the time dependent Hamiltonian is extremal. 

This works the same way in general relativity. We fix one of the variations in the symplectic form to be 
\beq
 \delta_Y \pi_V=2 (d-1)\alpha , \quad \delta_Y \bar \pi^{ab}=\delta_Y \sqrt{|h|}= \delta_Y \bar h_{ab}=0,
\eeq
so that \eqref{eq:sympgrav} becomes 
\beq
\label{eq:volumefromsympl}
\Omega (\delta_Y,\delta) =2 (d-1) \alpha \delta V.
\eeq
In terms of the induced metric and the extrinsic curvature, this transformation reads as
\bea
\label{eq:weyl}
\delta_Y h_{ab}=0, && \quad && \delta_Y K_{ab}= \alpha h^{ab}.
\eea
Again, one needs to make sure that this variation satisfies the constraints, since in general it is not possible to enforce $\delta_Y \sqrt{|h|}=0$, because $\sqrt{|h|}$ is a function of the remaining coordinates. The momentum constraint is automatically satisfied
\begin{equation}
\delta_Y(\nabla^a K_{a b}-\nabla_b K^{a}_{a} )=\nabla^a h_{a b}-\nabla_b h^{a}_{a}=0
\end{equation}
since the $\delta_Y \nabla_b$ is zero because it only depends on tangential derivatives of the metric and $\delta_Y K_{ab}\propto h_{ab}$  is covariantly constant. 
 The Hamiltonian constraint reads as
 \begin{equation}
\delta_Y( K_{a b} K^{a b}-(K_a^a)^2-R_d+2\Lambda)\propto 2 K-2 (d-1) K-0=2 (d-2) K
 \end{equation}
 where we used $\delta_Y K=\delta_Y h^{a b} K_{a b}+h^{a b} \delta_Y K_{a b}=\alpha h^{a}_{a}=\alpha (d-1)$. Therefore, we can only enforce $\delta_Y \sqrt{|h|}=0$, if $K=0$, i.e. the slice that we are considering is extremal.

How should we think about the deformation $\delta_Y$? York time is purely an internal time: it should move us inside the WdW patch, so it cannot possibly be physical. This intuition is reflected in the fact that, when one of the variations is a time translation (diffeomorphism), the Hamiltonian localizes to the boundary of the Cauchy slice and a York translation does not move the boundary time slice. So, York time translations do not seem to act physically in the boundary. As we will discuss later, this statement is a little subtle because if we put a bulk cutoff at a large radius, this shift will act in the boundary, but it will only affect the divergent terms. 
On the other hand, the new York transformation \eqref{eq:weyl} is not a diffeomorphism, as this transformation does not evolve the gauge invariant initial data $(\bar h, \bar \pi)$ in York time. Instead the new York transformation is ``copying" (instead of evolving) this initial data to a neighbouring slice. Note that we could have equivalently (up to boundary divergent terms) kept the slice fixed and shifted the phase space variables.

To gain a boundary understanding of the volume, we want to use the equality of the bulk and boundary symplectic forms in \eqref{eq:volumefromsympl}. To write a boundary formula, we need to understand the deformation $\delta_{Y} \gamma_{ab}=(\delta_{Y} \gamma_{ab}^-,\delta_{Y} \gamma_{ab}^+)$ of the background metric of the CFT that gives rise to the transformation \eqref{eq:weyl} upon solving the linearized bulk equations of motion.\footnote{As we will see, when there is backreacting matter in the background, the deformation $\delta_Y$ also involves sources for the operators dual to the matter.} In the following two sections, we will examine \eqref{eq:weyl} around the simplest possible backgrounds and determine the corresponding new York transformations in the boundary.

\section{The volume for deformations of the vacuum}

 \label{sec:vacuumdeformation}

\subsection{Bulk York transformation \label{Yorktrans}}

\begin{figure}[h!]
\centering
\includegraphics[width=0.45\textwidth]{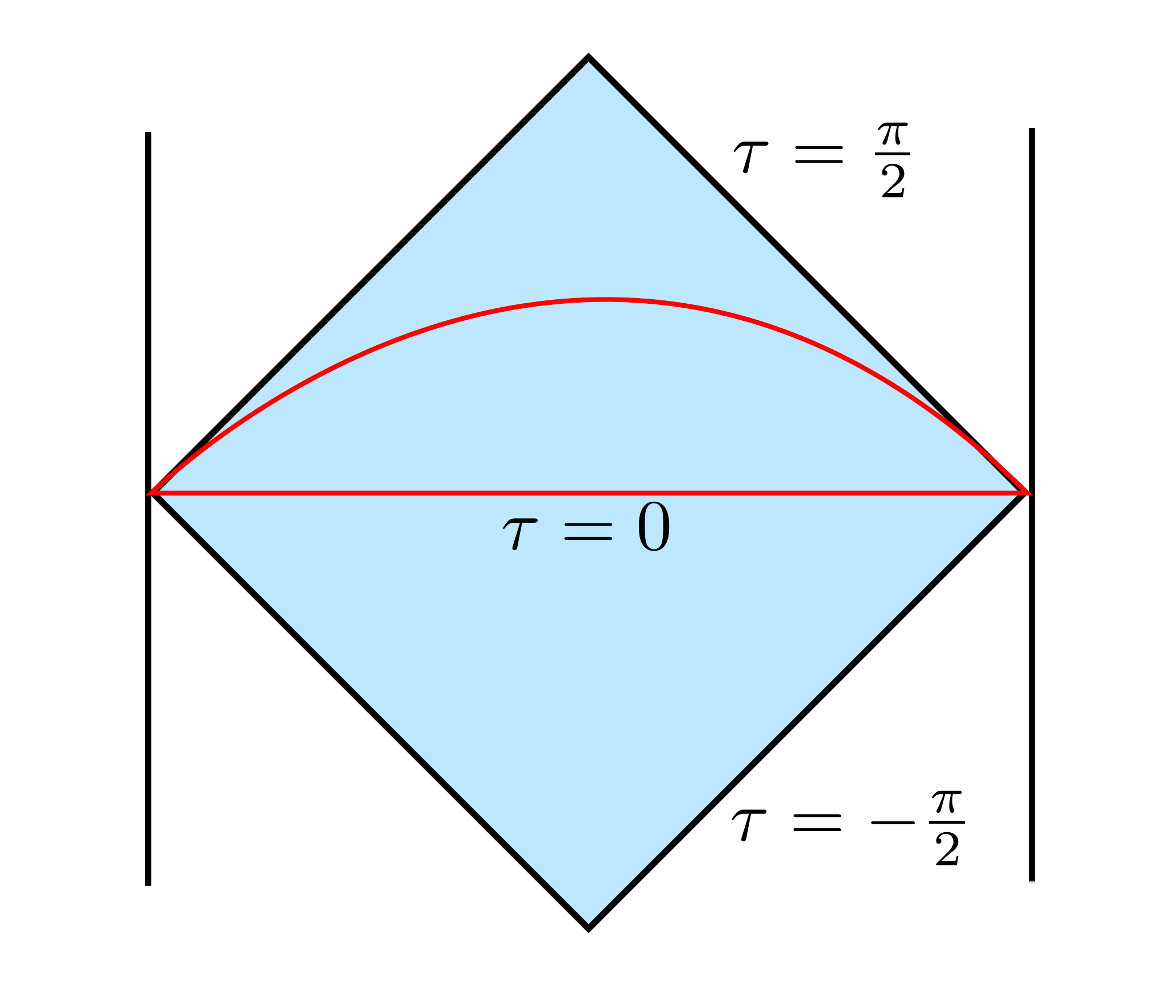} \includegraphics[width=0.45\textwidth]{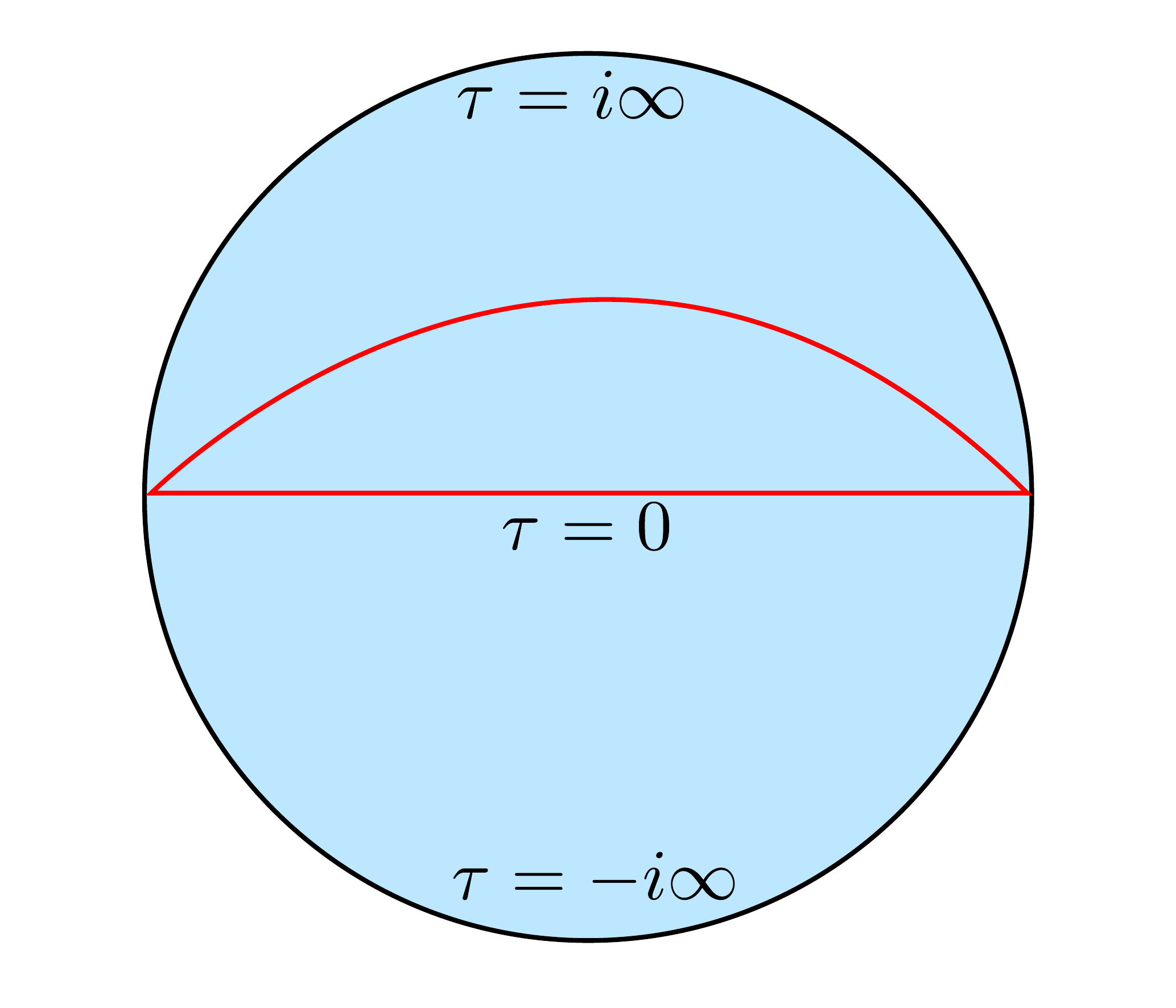}
\caption{We show the WdW patch in Lorentzian (left) and Euclidean AdS (right). }
\label{fig:2}
\end{figure}

Let us start by describing how the CMC slicing and the transformation \eqref{eq:weyl} works around empty AdS$_{d+1}$ space (or locally AdS space in $d=2$).
The CMC slicing is achieved by picking Wheeler-De Witt (WDW) coordinates\footnote{For some previous uses of this foliation, see \cite{Maldacena:2004rf,Maloney:2015ina}.}
\begin{equation}
\label{eq:WDWADS}
ds^2=-d\tau^2+ \cos^2 \tau d\Sigma_{d}^2
\end{equation}
where $d\Sigma_{d}^2=h_{ab}(x) dx^a dx^b$ is a $\tau$ independent Einstein metric satisfying $(R_{d})_{ab}=-(d-1)h_{ab}$. These coordinates cover the WDW patch in Lorentzian and the entire manifold in Euclidean. The Euclidean boundary is at $\tau = \pm i \infty$, see Fig. \ref{fig:2}. 

We summarize the relation of this coordinate system with the usual Poincar\'e coordinates in Appendix \ref{app:wdw}. This foliation of AdS has constant time slices with constant extrinsic curvature $K=- d \tan \tau$. Translation in York time $\pi_V$ is then given by 
\beq
\label{eq:vacuumweyl}
\partial_{\pi_V}=\frac{\cos^2 \tau}{2(d-1)} \partial_\tau,
\eeq
which on the maximal Cauchy slice, i.e. at $\tau=0$, is just $\frac{1}{2(d-1)}\partial_\tau$.
For these spacetimes, it is easy to see that the unconstrained phase space variables do not change under evolution with respect to York time, since clearly $\partial_\tau \bar{h}_{a b}=\partial_\tau \bar{\pi}_{a b}=0$. On the maximal slice we also have $\partial_\tau \sqrt{|h|}=0$, which means that in empty AdS, time evolution with respect to York time is the same as the new York transformation \eqref{eq:weyl}. This means that in the symplectic form \eqref{eq:volumefromsympl}, the transformation $\delta_Y$ is a diffeomorphism, and therefore any variation of the volume is a boundary term.\footnote{Since the volume is extremal, variations with respect to the shape of the anchoring slice always give boundary terms, however in the case of vacuum AdS, all metric variations give a boundary term.}

We can see explicitly how the volume is a boundary term by using that for AdS we have $(R_d)_{a b}=-(d-1) h_{a b}$ (here $(R_d)_{a b}$ is the Ricci tensor of the induced metric on the $\tau=0$ slice) and that, in the absence of matter, the Hamiltonian constraint imposes $\delta R_d(\tau=0)=0$. This gives
\begin{equation}
2 (d-1) \delta V= (d-1)\int_{\Sigma}\sqrt{|h|} h^{a b} \delta h_{a b}=-\int_{\Sigma} \sqrt{|h|}(R_d)^{a b} \delta h_{a b}=-\int_{\Sigma}\sqrt{|h|} h^{a b} \delta (R_d)_{a b}
\end{equation}
which is a boundary term at $t=0$ because of the usual Palatini identity. In fact, we can write it solely in terms of curvature invariants associated to $\partial \Sigma$ as
\beq
\label{eq:volumbndyterm}
2 (d-1) \delta V=\int_{\partial \Sigma}\sqrt{|\hat h|}(2 \delta \hat K-\hat K^{ij}\delta \hat h_{ij}),
\eeq
where the hat refers to quantities associated to the codimension two surface $\partial \Sigma$.
A similar observation was made before in \cite{Fu:2018kcp} for $d=2$.  In fact, around flat space, using the general divergence structure of the volume \cite{Carmi:2016wjl}, one sees that the only divergence after integrating properly by parts is this leading $\delta \gamma^{i}_{i}$ divergence. The fact that there is no finite contribution can be easily checked using Euclidean HKLL as follows. We can write a linarized deformation of Poincar\'e AdS
\beq
ds^2=\frac{1}{z^2}\Big[ dz^2+dt^2+dx^i dx^i + \delta H_{ab}(z,t,x)dx^a dx^b\Big],
\eeq
for which the change in the volume is
\bea
\delta V & = \frac{1}{2} \int_\epsilon^\infty z^{-d} \int d^{d-1} x \delta H_{ii}(z,t=0,x^j) \\
&= \frac{1}{2} \int_\epsilon^\infty z^{-d} \int_{-\infty}^\infty d \omega \delta H_{ii}(z,\omega,k_j=0),
\eea
where the trace $\delta H_{ii}$ is over the $d-1$ spatial indices orthogonal to the FG coordinate $z$. The Fourier transform $\delta H_{ab}(z,\omega,k_i)$ can be expressed with the Fourier transform of the boundary metric deformation in a simple way by solving linearized EOM with boundary condition $\delta H_{ab}(z=0)=\delta \gamma_{ab}$, see \cite{Marolf:2017kvq} (e.q. (69) in particular).
Here we only need the spatial trace for $k_j=0$ which is very simple
\beq
\delta H_{ii}(z,\omega,k_j=0)=\delta \gamma_{ii}(\omega,k_j=0),
\eeq
in particular, all the $z$ dependence cancels.
This leads to the change in the volume
\beq
\label{eq:vacuumvolume}
\delta V = \frac{1}{2(d-1)\epsilon^{d-1}} \int d^{d-1}x \delta \gamma_{ii}(t=0,x),
\eeq
which is a pure divergence, and only contains the leading term in \cite{Carmi:2016wjl}. As mentioned earlier, the absence of any finite contribution can be understood from the fact that evolution in York time moves inside the WdW patch and hence it is only non-trivial at the boundary because of the presence of the cutoff. Moreover, in the class of ``nice states" that we want to consider, we always switch off boundary sources at the $t=0$ surface, in which case \eqref{eq:vacuumvolume} gives zero. The fact that around the vacuum $\delta V$ is a pure divergence will be important later when we try to make a connection to the complexity equals volume conjecture.

\subsection{Boundary York transformation}

Let us turn to the boundary story and read the deformation of the boundary metric that induces \eqref{eq:weyl}. In WDW coordinates \eqref{eq:WDWADS}, the deformation is just a shift in time, which in the metric amounts to the replacement $\cos^2 \tau \mapsto \cos^2 \tau(1+  2\alpha \tan \tau )$. We can read the metric deformation on the top and the bottom Euclidean boundaries by sending $\tau \rightarrow \pm i \infty$ respectively, which leads to $\delta^{\pm}_Y \gamma_{ab} = \pm 2 i \alpha \gamma_{ab}$\footnote{We note that this is the same as the complex structure \eqref{eq:J} acting on a constant Weyl rescaling.}, or in terms of the total source
\bea
\label{eq:signdeformation}
\delta_Y \gamma_{a b} & = 2 i \alpha \text{sign} (\tau_E) \gamma_{a b}
\eea
Since this is in a hyperbolic conformal frame, it is instructive to read the deformation also in Poincar\'e coordinates. Using the coordinate changes summarized in Appendix \ref{app:wdw}, we have that the deformation is generated by the (Euclidean) vector flow
\beq
\label{eq:poinvacweyl}
\xi_Y=i\partial_{\tau_E} = \frac{ -i t_E z}{\sqrt{t_E^2+z^2}} \partial_z + \frac{i z^2}{\sqrt{t_E^2+z^2}}\partial_{t_E},
\eeq
where $t_E$, $z$ are the usual Euclidean coordinates in Poincar\'e. Near the boundary, this approaches $-i\text{sign}t_E z \partial_z$. It induces a rescaling of the boundary metric by a Weyl factor $(1-2 i \alpha \text{sign}t_E)$, resulting in \eqref{eq:signdeformation} 

The complete source deformation entering the boundary partition function is therefore proportional to a sign function, and it is discontinuous, which is an ill-defined situation, and needs UV regularization. From the boundary point of view, a natural regularization would be to switch off the sources in a small buffer zone around $t_E=0$, while keeping the deformation \eqref{eq:signdeformation} outside of this. This would give, using that $\langle T_{ab}\rangle =0$ in the background,
\beq
\label{eq:vacsymplbndy}
\Omega(\delta_Y,\delta) \propto \int_{t_E<-\epsilon} \delta^+ \langle T^a_a \rangle +\int_{t_E>\epsilon} \delta^- \langle T^a_a \rangle =0. 
\eeq
The reason that this expression is zero is simply that the trace anomaly $\langle T^a_a \rangle$ is always a local function of the sources, and here we always integrate the change in the trace anomaly on the opposite side compared to where there is a nontrivial variation. Note that this expression is indeed only well-defined if $\epsilon>0$, since otherwise contact terms in the $\langle T^a_a T_{cd} \rangle$ correlator contribute.\footnote{In two dimensions, the contact term is universal, but gives a divergent contribution. In higher dimensions, the contact terms depend on the type of counter terms that we add.} So how does this compare to the bulk answer for the change in the volume \eqref{eq:vacuumvolume}, which can be non-zero? The answer is that the bulk uses a different UV regulator. The deformation \eqref{eq:poinvacweyl} on a finite cutoff surface $z=\epsilon$ gives rise to the change in the boundary background metric\footnote{ Note that a different situation where a $\text{sign}(t)$ is regulated by the bulk has been observed before in Janus geometries \cite{Bak:2007jm}.} 
\beq
ds^2 = \left(1-2i\alpha \frac{ t_E^3}{(t_E^2+\epsilon^2)^{\frac{3}{2}}} \right)dt_E^2+ \left(1-2i\alpha \frac{ t_E}{\sqrt{t_E^2+\epsilon^2}} \right)dx^2.
\eeq
This is an anisotropic regularization of the sign multiplier, and one can check that in the bulk symplectic form, \eqref{eq:vacuumvolume} comes precisely from this anisotropy.\footnote{The bulk symplectic form is obviously conserved and we can push it near the boundary. There, the gravitational momenta can be exchanged with the holographic stress tensor and the induced metric with the boundary background metric. This is because the rescalings cancel between the products in the symplectic form, while the counterterms to the stress tensor cancel between the antisymmetrization.} However, we do not have a clear boundary understanding of this bulk regulator, since the diffeomorphism \eqref{eq:poinvacweyl} introduces $dz dt$ terms, under which the holographic stress tensor transforms in a nonstandard way. 
Indeed, a rescaling of $z$ normally amounts to a change of conformal frame in the boundary, but in the present case the Weyl factor is $z$ dependent. This dependence is ``mild enough" so that when $t_E\gg z\approx \epsilon$ we get a boundary Weyl transformation, but it cannot be neglected when $t_E \approx z$. It would be interesting to understand this better, but for now, the main lesson of this story is that when the source deformations are not switched off at $t_E=0$, divergences appear in both the bulk and boundary symplectic forms, and these require a choice of UV regulator. For states and their deformation for which $\delta V$ is finite, we expect no regulator dependent ambiguities. We will discuss in the next section such an example.

As a final comment, note that because the new York deformation for the vacuum is a diffeomorphism, we could have integrated by parts directly at the level of the symplectic form, using the formalism of \cite{Iyer:1994ys}. Of course, the boundary term that we get is the same as \eqref{eq:volumbndyterm}. More concretely, for a given boundary value of the vector field, $\xi$,  Wald's boundary term normally has two terms: one which depends on $\nabla \xi$, which is denoted $\delta Q[\nabla \xi]$, and one that depends only on $\xi$: $\xi.\Theta(\delta g)$. Using the FG expansion, if we have a vector field tangent to the  AdS boundary at $z=\epsilon$, there is no contribution from $\delta Q$ and $\xi.\Theta$ is identified with the boundary (Brown-York) stress tensor, see for example \cite{Faulkner:2013ica}. However, this becomes subtle in WdW coordinates and the interpretation of \eqref{eq:volumbndyterm} in terms of $\delta \langle T_{\tau \tau} \rangle$ in the boundary hyperbolic geometry asymptotic to the WdW patch is not straightforward.  In any case, we believe there should be some way to understand \eqref{eq:vacuumvolume} from a boundary calculation.

\subsection{An example with finite contributions: scalar condensates}
\label{sec:scalar}

We have seen that it is not possible to have a finite change in the volume around vacuum AdS by looking at linearized deformations of only the background metric of the CFT. However, it is possible to have a finite variation, if we slightly shift the background, e.g. by turning on sources for some other operator than the stress tensor. 

In this section, we will consider the case where we turn on a scalar operator with source $ \lambda=(\lambda^-,\lambda^+)$, and we treat the background source perturbatively, to lowest order where the backreaction appears, i.e. $\lambda^2$. 

We want to compare the change in the volume with the symplectic form, and to read the boundary version of the York deformation \eqref{eq:weyl}. Let us first count orders of $ \lambda$. The change in the volume will be kept to order $ \lambda^2$, and the variation of this is order $ \lambda \delta  \lambda$. Since this is also the leading order variation in the bulk metric, to match this with the gravitational symplectic form, we only need to keep the metric background and the York deformation to order $ \lambda^0$, i.e. we can use the vacuum York deformation \eqref{eq:vacuumweyl}. Note that this appears to be a diffeomorphism, and it indeed is one around the vacuum. However, we now have a scalar condensate in the bulk with a field profile which is order $ \lambda$. Thus if we regard \eqref{eq:vacuumweyl} as a diffeo, it must act on not only on the metric but on this condensate as well and this will also contribute to the scalar part of the bulk symplectic form at order $ \lambda \delta \lambda$. Therefore, acting with \eqref{eq:vacuumweyl} on the entire background is not equal to \eqref{eq:weyl} with all other deformations being zero, in particular, it does not give the volume. In order to recover \eqref{eq:weyl} and the volume, we need to cancel the change in the scalar initial data, which can be done by writing (in the WDW coordinates \eqref{eq:WDWADS})
\bea
\delta_Y g_{ab} &= \alpha \mathcal{L}_{\partial_\tau} g_{ab}\\
\delta_Y \phi &= \alpha \partial_\tau \phi+ \delta' \phi,
\eea
and fixing $\delta'$ by requiring that on the initial value surface $\tau=0$, we have for the initial data $(\varphi,\pi)$
\be
\delta_Y \varphi = \delta_Y \pi =0 \,.
\ee
Using that $\delta_Y$ must solve the linearized scalar-metric equations of motion, we conclude that $\delta' \phi$ must solve the scalar equation of motion in vacuum AdS with initial data
\bea
\label{eq:Yorkprime}
\delta ' \varphi &=-\alpha \partial_\tau \phi|_{\tau=0}=0, && \quad && \delta ' \pi &=-\alpha \partial_\tau^2 \phi|_{\tau=0}=0.
\eea
In this way, the York deformation  in the presence of scalar sources will be the diffeomorphism \eqref{eq:vacuumweyl}, combined with a deformation $\delta'$ that only affects the scalar sector. Note that the backreaction of $\delta'$ to the metric can be neglected at the order at which we are working.

Let us examine the gravitational and the scalar symplectic forms when one of the deformations is $\partial_\tau$. For the scalar, it is clear that we must get the variation of the bulk Hamiltonian, i.e.
\beq
\label{eq:scalarsympl}
\Omega_{\rm scalar} = \int_\Sigma (\delta \pi \delta'\varphi-\delta' \pi \delta \varphi) = -\int_\Sigma \sqrt{|h|} \delta T^{\rm scalar}_{\tau\tau},
\eeq
where $T^{\rm scalar}$ is the bulk stress tensor for the scalar. For the gravitational part of the symplectic form, let us assume that the scalar background is $Z_2$ symmetric, i.e. the half sided source $\lambda^-$ is real. In this case, we have $K_{ab}=0$ in the background, so taking a variation of the Hamiltonian constraint in the presence of the matter stress tensor gives 
\beq
\delta R_d = \delta T^{\rm scalar}_{\tau \tau}
\eeq
We can therefore write the leading $ \lambda \delta \lambda$ order variation of the volume as
\bea
\label{eq:scalarvolume1}
2(d-1)\delta V &= (d-1)\int_\Sigma \sqrt{|h|}h^{ab}\delta h_{ab} = 
-\int_\Sigma \sqrt{|h|}(R_d)^{ab}\delta h_{ab} \\
&=
\int_\Sigma \sqrt{|h|} \delta T^{\rm scalar}_{\tau \tau}+\text{boundary term}
\eea
where in the second equality, we used the Hamiltonian constraint and integrated by parts as in \eqref{eq:volumbndyterm}. We will focus on a scalar source which vanishes fast enough at $t=0$ so that there is no contribution from the boundary term.\footnote{Putting the diffeomorphism $\partial_\tau$ in the full symplectic form produces only this boundary term by adding up \eqref{eq:scalarsympl} and \eqref{eq:scalarvolume1}. It is the same as in \eqref{eq:volumbndyterm}, and from the boundary point of view, it can depend only on a finite number of derivatives of the sources at $t=0$.} 
 We see that in this case the change in the volume is controlled only by the change in the scalar energy, and can be obtained by a symplectic pairing only in the scalar sector, \eqref{eq:scalarsympl}.

We see something interesting: we can either think about $\delta_Y$ as having contribution only from the gravitational part of the symplectic form, or as having contribution from a diffeo, for which the gravitational and scalar parts cancel, and the volume in the end comes entirely from the scalar sector. It turns out that this latter is a more natural interpretation from the boundary point of view. In the boundary, the effect of the diffeo \eqref{eq:vacuumweyl} is to do a Weyl transformation on all the sources, $ \lambda \mapsto \Phi^{d-\Delta}  \lambda$, $\gamma_{ab} \mapsto \Phi^2  \gamma_{ab}$, with $\Phi=(1+i2\alpha \text{sign}t_E)$. But this transformation is controlled by the trace Ward identity, in particular, we get \eqref{eq:vacsymplbndy} but $\langle T^a_a\rangle$ is replaced by $\langle T^a_a\rangle +(d-\Delta)\lambda \langle O \rangle$ which is a local anomaly again, and the boundary symplectic form vanishes for the same reason as explained after \eqref{eq:vacsymplbndy}. This way, the variation of the volume entirely comes from the scalar sector in the boundary, namely from the $\delta'$ deformation.  We note that this fact is more of a feature than a bug, if we want to think that the volume measures some kind of distance between pure states as is the case for the complexity = volume conjecture. This is because we do not use the background metric at all to define the path integral state $|\lambda \rangle$ and there is no reason to expect that creating this state in some ``optimal" way would refer to the stress tensor sector.

\subsubsection*{Large $h$ approximation}

Let us explore this in more detail via an example. We want to turn on some source $\lambda$ for a scalar operator to create the state $|\lambda \rangle$, and we want to use the standard dictionary to obtain the corresponding bulk solution to leading order in the source, as in \cite{Marolf:2017kvq}. To derive the new York transformation, it will be convenient to work in WDW coordinates \eqref{eq:WDWADS} instead of Poincar\'e. In this description, the boundary CFT naturally lives in a hyperbolic conformal frame, where the CFT background metric is the same as the metric of the $\tau=0$ slice in AdS. We will also resctrict to $d=2$ for simplicity. The boundary to bulk propagator in these coordinates is
\beq
\label{eq:wdwbulktobndy}
K_E(\tau_E,u,x|u_0,x_0) = c_h \left(\frac{1}{\cosh \tau_E}\frac{2 u u_0}{(x-x_0)^2+u^2+u_0^2+2 s u u_0 \tanh \tau_E} \right)^{2h},
\eeq
where $u_0$ and $x_0$ are the boundary coordinates and we have picked the time slice part of the WdW metric to be in Poincar\'e coordinates, as in \eqref{WdWmetric}. The parameter $s$ is $+1$ when the boundary point is taken to be on the top boundary and $-1$ when it is on the bottom boundary (see right of Fig. \ref{fig:2}). The coefficient $c_h$ is fixed such that $e^{-2h\tau_E}K_E$ approaches a delta function as $\tau_E \rightarrow \infty$.\footnote{Explicitly, $c_h=\frac{\Gamma(2h)}{4\pi \Gamma(2h-1)}$.} We want to focus on sources independent of $x_0$, so that we only have to deal with a $u_0$ integral. We can integrate out $x_0$ easily from \eqref{eq:wdwbulktobndy} by introducing a Schwinger parameter to deal with the power function, the result is 
\bea
\label{eq:intwdwbulktobndy}
\int dx_0 & K_E(\tau_E,u,x|u_0 ,x_0) \\ &= c_h \frac{\sqrt{2\pi u u_0}\Gamma(2h-\frac{1}{2})}{\Gamma(2h)}(\cosh\tau_E )^{-2h}\left( \frac{u}{2u_0}+\frac{u_0}{2u} + s \tanh \tau_E \right)^{\frac{1}{2}-2h}.
\eea
The bulk solution is given by integrating this against a source $\lambda^-(u_0)$ on the bottom, and a source $\lambda^+(u_0)$ on the top.
To proceed, we assume that $h$ is large (but $O(N^0)$), in which case \eqref{eq:intwdwbulktobndy} becomes highly peaked as a function of $u_0$ and we can do a saddle point approximation to the $u_0$ integral, resulting in\footnote{This expression is valid for any $\tau_E$ because the coefficients of $h^{-1}$ corrections are bounded functions of $\tau_E$.}
\bea
\label{eq:largehsol}
\phi(\tau_E,u)
&=\frac{\pi c_h}{h \cosh \tau_E}\Big( e^{(1-2h) \tau_E} \lambda^-(u)+e^{(2h-1) \tau_E} \lambda^+(u) \Big)\Big(1+O(h^{-1} )\Big),
\eea
which is a very simple relation between the bulk field $\phi$ and the boundary source. Note that it is easy to see at this point that we need $c_h=\frac{h}{2\pi}$ to leading order in large $h$ for the propagator to have the correct delta function normalization. The relation between Lorentzian initial data and the sources is equally simple
\bea
\label{eq:largehdata}
\varphi(u)& = -\frac{\pi }{h} c_h[\lambda^-(u)+\lambda^+(u)]\Big( 1+ O(h^{-1}) \Big),\\
\pi(u) &= 2 \pi i c_h[\lambda^-(u)-\lambda^+(u)]\Big( 1+ O(h^{-1}) \Big).
\eea
Notice that this is real initial data, as it should be, since $\lambda^\mp$ are complex conjugates. The transformation of initial data required to get the volume, defined in \eqref{eq:Yorkprime}, are obtained from the first and second $\tau$ derivative of \eqref{eq:largehsol} at $\tau=0$
\bea
\label{eq:scalarweyl}
\delta' \varphi(u) & =i\frac{2h-1}{h} \pi c_h[\lambda^-(u)-\lambda^+(u)],\\
\delta' \pi(u) &= - 4\pi (h-1)c_h[\lambda^-(u)+\lambda^+(u)].
\eea
The scalar symplectic form between this initial data and a variation of \eqref{eq:largehdata} gives
\beq
\label{eq:scalarvolume}
\frac{1}{u^2} (\delta \pi \delta' \varphi - \delta' \pi \delta \varphi) = c_h^2\frac{8\pi^2}{u^2}[\delta \lambda^-(u) \lambda^+(u)+\delta \lambda^+(u) \lambda^-(u)] + O(h^{-1}),
\eeq
which is indeed given by the variation of the energy density $\sqrt{|h|}\delta T_{\tau \tau}^{\rm scalar}$, as discussed before. 

Now let us read the boundary version of $\delta'$. Because it is easy to invert \eqref{eq:largehdata}, we can read off the sources corresponding to the deformation \eqref{eq:scalarweyl}:
\bea
\label{eq:boundaryscalarweyl}
\delta' \lambda^-(u)& = i 2 h \lambda^-(u)\big(1+O(h^{-1})\big), && \delta' \lambda^+(u)& = -i 2 h \lambda^+(u)\big(1+O(h^{-1})\big),
\eea
i.e. to leading order in $h$, it is just a sign deformation again, but now for only the scalar sources. Note that at subleading orders $\delta' \lambda^-$ will contain both $\lambda^-$ and $\lambda^+$ contributions. This is therefore an explicit example of a boundary deformation that gives a finite variation of the volume via formula \eqref{eq:volumefromsympl}. In appendix \ref{app:scalarcond}, we explicitly verify that \eqref{eq:scalarvolume} is recovered using the boundary symplectic form with \eqref{eq:boundaryscalarweyl} inserted in one of the slots, and that this agrees with the variation of the backreacted volume in the background \eqref{eq:largehsol}.

\section{The volume and the late time thermofield double state}
\label{sec:inftfd}

From the point of view of the volume of the maximal Cauchy slice, a particularly interesting state is the thermofield double state under time evolution. This state is dual to the eternal AdS-Schwarzschild black hole with an initial data surface that is anchored at $t_L=t_R=\cal{T}$.\footnote{Our convention is that $H_L-H_R$ annihilates the state and $t_L=t_R=\cal{T}$ corresponds to time evolution with $H_L+H_R$.} 
In this state, the dependence of volume with time has been extensively studied in \cite{Stanford:2014jda,Carmi:2017jqz,Kim:2017qrq}. In summary, the rough behavior is that the volume grows quadraticaly with time at early times and linearly at late times, and the rate is given by the energy $M$ of the state:
 \begin{equation}
 \label{eq:TFDvolume}
 2 (d-1) V({\cal T}\gg 1)=2 M \cal{T}+...
 \end{equation}
 where the $...$ are terms that do not grow with time. The reason why this behavior sparkled much recent interest is that it captures the growth of the interior of the wormhole even at very late times, long after probes of thermalization saturate. Therefore it is desirable to identify what this growth measures for the state of the dual CFT.
 
 \subsection{New York transformation}
 
For generic times, we expect that it is very hard to understand the York transformation \eqref{eq:weyl}. However, the maximal volume slice becomes simple at very late times \cite{Stanford:2014jda} and in this case we will be able to solve the problem. We can think of the WDW patch of an infinite time Cauchy slice as the whole interior region (the region between the horizon and the singularity). In the coordinates of \cite{Hartman:2013qma}, we can write the metric in the interior for a planar black hole of temperature $\beta^{-1}$ as 
\begin{equation}
ds^2=\left (  \frac{2}{d} \right)^2 [-d\kappa^2+\left(\frac{2 \pi}{\beta} \right)^2(\cos \kappa)^{\frac{4}{d}} (dx_{d-1}^2+ \tan^2 \kappa dt^2 ) ] \label{interior}
\end{equation} 

 \begin{figure}[h!]
\centering
\includegraphics[width=0.45\textwidth]{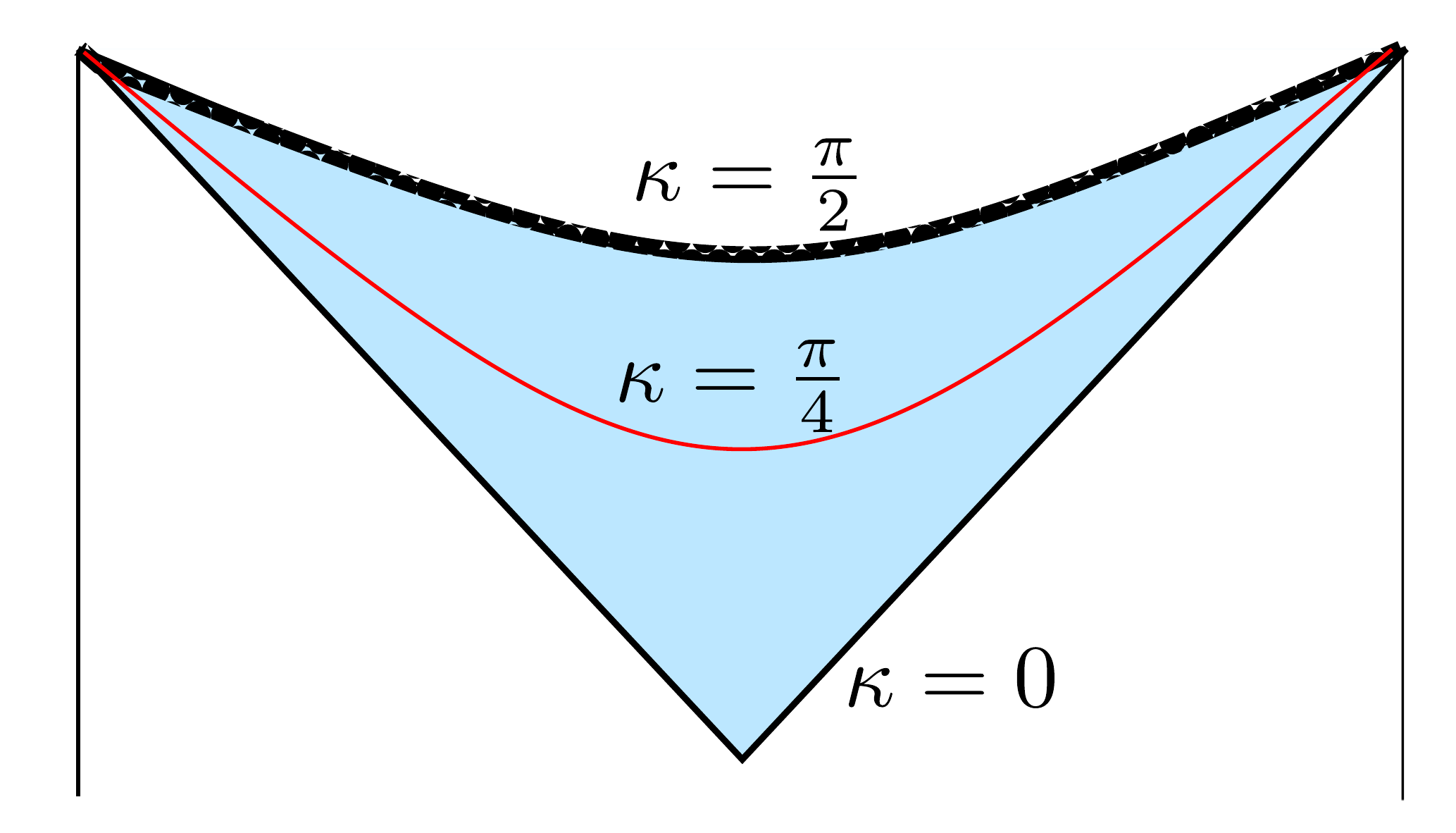}
\caption{The top part of the Penrose diagram of the eternal wormhole with the location of the horizon, the maximal infinite time Cauchy slice and the singularity highlighted in the coordinates \eqref{interior}. The blue region is the top interior, and also the WDW patch of the red slice.}
\label{fig:3}
\end{figure}

This is a perfectly fine metric, with $t$ being a spacelike coordinate. Constant $\kappa$ surfaces are spacelike surfaces and the boundary of this geometry at ${t}=\pm \infty$ corresponds to the left/right  $t_L=t_R=\infty$ boundary Cauchy slices. Some interesting surfaces are located at $\kappa=0$ (the horizon), $\kappa=\pi/2$ (the singularity) and $\kappa=\pi/4$ (the extremal surface), see Fig. \ref{fig:3}. Because of the symmetries, constant $\kappa$ surfaces correspond to constant extrinsic curvature surfaces and thus we can think of $\kappa$ as York time.

The surface $\kappa=\frac{\pi}{4}$ in this geometry is, to our knowledge,  the simplest case of an extremal but not $Z_2$ symmetric surface. This will allow us to understand rather explicitly how to think about the new York transformation \eqref{eq:weyl}. We can get the explicit deformation by turning on small deformations $\delta g_{{t} {t}}=\alpha g(\kappa), \delta g_{x_i x_i}=\alpha f(\kappa)$ in \eqref{interior}. Asking for $g(\pi/4)=f(\pi/4)=0$, we get:
\begin{eqnarray}
f(\kappa)= \left(\frac{2 \pi}{ \beta}\right)^2 \left (\frac{2}{d} \right)^3 \cos^{4/d} \kappa (\tan \kappa-1) \nonumber \\
g(\kappa)=\frac{1}{2} f(\kappa) \tan \kappa (d-(d-2) \tan \kappa)
\end{eqnarray}
This deformation can also be obtained by applying the diffeo
\bea
\label{eq:infdiffeo}
\kappa &\mapsto \kappa-\alpha/2 \\
t &\mapsto(1+ \frac{d-1}{d} \alpha )t \\
x^j & \mapsto (1-\frac{1}{d} \alpha )x^j
\eea
to \eqref{interior}. We can understand this as obtaining \eqref{eq:weyl} by evolving a bit in York time, and then cancelling $\partial_\kappa\bar h_{ab}$ and $\partial_\kappa\bar \pi^{ab}$ by a coordinate change inside a fixed $\kappa$ slice. Note that while this is a diffeomorphism, it is different from the case of vacuum AdS since it is not a pure York time translation.\footnote{In the vacuum, York translations and new York transformations were equivalent because there was a time translation symmetry.} We will see that it acts non-trivially at the boundary time slice, which will allow us to obtain cutoff independent variations in the volume.

Our next aim will be to read the deformation of the boundary background metric that gives rise to \eqref{eq:infdiffeo} and use this to reproduce some simple variations of the late time volume \eqref{eq:TFDvolume} by a boundary calculation. We will focus on obtaining $\partial_{{\cal T}} V$ and $\partial_\beta V$ as two simple examples.\footnote{Note that $M \propto \beta^{-d}$ on dimensional grounds.} Before moving on, it is instructive to check how the local variation of the induced metric on the $\kappa=\pi/4$ slice gives rise to these derivatives directly in the bulk. 

Let us start with the $\beta$ derivative, for which we have $\partial_\beta h_{ab}=-2 \beta^{-1} h_{ab}$ for the induced metric of constant $\kappa$ slices. Therefore, the derivative of the volume of the $\kappa=\pi/4$ slice is
\bea
\label{eq:betaderivative}
2 (d-1) \partial_{\beta} V(t={\cal T}) &=2 (d-1)\int \sqrt{|h|} h^{a b} \partial_{\beta} h_{a b} \\ &=-4 d (d-1)/\beta V({\cal T})= -\frac{2 d M}{\beta} 2 {\cal T}+...
\eea
where we used the usual expression $V=\frac{M}{2 (d-1)} 2 \cal{T}$ and the $\cal{T}$ is obtained by putting a sharp cutoff in the $t$ integral. The $...$ refer to contributions which do not grow with $\cal{T}$. 

Now let us discuss $\partial_{\cal{T}} V$. We take $\delta_t h_{ab}$ to be the result of a diffeomorphism $\xi_t=f(t)\partial_t$. Since the volume is extremal, this variation must be a boundary term and we expect that the result only depends on the properties of $f(t)$ around the corners of the Penrose diagram Fig. \ref{fig:3}. Indeed, we have $\delta_t h_{ab}=2f'(t) h_{tt} \delta_{ta}\delta_{tb}$ which gives 
\bea
\label{eq:Tderivative}
2 (d-1) \partial_{\cal{T}} V= (d-1)\int \sqrt{|h|}h^{ij}\delta_t h_{ij} &=(d-1)\left( \frac{4 \pi}{d \beta} \right)^d \text{Vol}_{d-1}\int_{-\infty}^\infty dt f'(t) dt \\ &= M[f(\infty)-f(-\infty)].
\eea
The factor $[f(\infty)-f(-\infty)]$ gives $0$ if the diffeo approaches the Killing time $\partial_t$ in the corners, which corresponds to evolution with $H_L-H_R$, while $2$ if it approaches ``both forward" time, which corresponds to evolution with $H_L+H_R$.

\subsection{Pushing to the boundary}

To push the symplectic form to the boundary, we want to think about the time evolved thermofield double as being created by a CFT path integral. We recall that without the time evolution, the thermofield double is created by performing the path integral on half of the thermal cylinder, thus obtaining a state of two copies of the CFT. We want to think about the path integral for the time evolved state as attaching a Lorentzian part to the Euclidean path integral on the half cylinder. This way, expectation values of operators in the time evolved state are calculated by path integrating on a time fold, as is usual in the Schwinger-Keldysh formalism. The variations entering the boundary symplectic form therefore naturally live on half of this boundary time fold.

We can reach the exterior region from the bulk metric \eqref{interior} by analytically continuing via the horizon, by setting $\rho=-i\kappa$ \cite{Hartman:2013qma}. We can then reach the boundary by approaching $\rho \rightarrow \infty$, and our location on the time fold is determined by whether we continue the coordinate $t$ along the gluing surface of the Euclidean and Lorentzian segments at $t=0$. We can easily read the deformation of the boundary background metric at $\kappa\rightarrow i\infty$ and it is
\bea
\label{eq:inftimeyork}
\delta_Y \gamma_{tt}&=\alpha \frac{2}{d}  (d-1-i ), &&\quad && \delta_Y \gamma_{x_k x_l}&=-\alpha \frac{2}{d} (1+i) \delta_{kl} .
\eea
We can think about this as being induced by the $(t,x)$ diffeo in \eqref{eq:infdiffeo}, plus an imaginary Weyl rescaling coming from the shift in $\kappa$. Note that only this latter changes the boundary scale, since $\delta_Y \sqrt{\gamma}=-i \alpha$

Now let us discuss how to recover \eqref{eq:betaderivative} and \eqref{eq:Tderivative} directly from the boundary.
In the bulk, it is clear that we can write both \eqref{eq:betaderivative} and \eqref{eq:Tderivative} on any constant $\kappa$ slice by pairing $(\partial_\beta h_{ab},\partial_\beta \pi^{ab})$ and $(\delta_t h_{ab},\delta_t \pi^{ab})$ respectively with the deformations of $h_{ab}$ and $\pi^{ab}$ coming from applying the diffeo \eqref{eq:infdiffeo}. The fact that this symplectic flux is $\kappa$ independent can easily be checked explicitly and therefore we can just send it to the boundary by sending $\kappa \rightarrow i\infty$, and read the correct stress tensor deformations using the extrapolate dictionary. 
In order to understand the flux we are deforming by doing this, it is probably better to think of our geometry as an Euclidean complex geometry, in which we are deforming continuously from $\kappa=\pi/4$ to $\kappa=i \infty$ (note that this contains both boundaries and the whole time contour). From the Lorentzian perspective, these constant $\kappa$ surfaces are not spacelike once we exit the interior. 

For the $\beta$ derivative we have that $\partial_{\beta} \gamma_{a b}=0$ for the boundary metric, while the background stress tensor $\langle T_{ab} \rangle$ is completely determined by the total energy $M$, the fact that it is traceless, and that there is translational invariance in the spatial directions. Using this, we get the boundary symplectic pairing
\begin{align}
\label{eq:betaderivativesympl}
\Omega(\delta_{Y} \gamma,\delta_{\beta} \gamma)& = -i\int_{-\beta/2-i {\cal T}}^{\beta/2+i {\cal T}}  dt \int dx \partial_{\beta}\langle T_{a b}\rangle \delta_Y \gamma^{a b} \\ &= i \frac{d M}{\beta} \int_{-\beta/2-i {\cal T}}^{\beta/2+i {\cal T}} (\delta_Y \gamma^{00}-\frac{1}{d-1}\delta_Y \gamma^{i}_{i} ) \nonumber \\ &=  \frac{-2 d M}{\beta} (2 {\cal T} +i \beta) &
\end{align}
This agrees with \eqref{eq:betaderivative} as ${\cal T} \rightarrow \infty$. Notice that $\delta_Y \langle T^{ab} \rangle$ did not contribute to this calculation. In order to get a more precise matching at finite ${\cal T}$, one would have to deal with the fact that the extremal surface is not a constant $\kappa$ surface and understanding the York deformation becomes much harder. 

For the ${\cal T}$ derivative, we have $\delta_t \gamma_{ab} = 2 f'(t)\delta_{at}\delta_{bt}$ for the boundary metric. We can read the stress tensor deformation from the bulk. Under diffeos inside constant $\kappa$ slices, it is clear that the holographic stress tensor just transforms as a two index tensor. Under the shift in the bulk radial coordinate $\kappa$, the renormalized holographic stress tensor undergoes a usual Weyl transformation $\langle T_{ab}\rangle \mapsto \Phi^{2-d}\langle T_{ab}\rangle$ \cite{deHaro:2000vlm}.
\footnote{It might seem confusing that a bulk radial shift has different effect on the boundary stress tensor than a coordinate rescaling, since these two are indistinguishable at the level of the change in the background metric components \eqref{eq:inftimeyork}. From the bulk point of view, these are different because only the radial shift involves a variation in the pull-back coming from changing the location of the boundary surface. 
In the boundary, on top of changing the metric, a Weyl transformation also changes the conformal class. This gives a different transformation for the stress tensor. } In summary, we can obtain the stress tensor variations from the usual CFT transformation formula
\beq
T_{ab} \mapsto \Phi^{2-d} \frac{\partial x^c}{\partial y^a}\frac{\partial x^d}{\partial y^b}T_{cd},
\eeq
under a combined diffeo $x\mapsto y(x)$ and Weyl transformation $\gamma_{ab} \rightarrow \Phi^2 \gamma_{ab}$. For $\delta_t$, the Weyl factor is $\Phi=1$, while the diffeomorphism is $t\mapsto t+f(t)$, while for the $\delta_Y$, the Weyl factor is $\Phi = 1+i\frac{\alpha}{d}$ and the diffeo is given by the last two lines in \eqref{eq:infdiffeo}. The explicit variations coming from this read as
\bea
\delta_t [\sqrt{\gamma}\langle T^{tt}\rangle] &=-f'(t)\langle T^{tt}\rangle, \quad \quad
\delta_t [\sqrt{\gamma}\langle T^{x_kx_k}\rangle] =f'(t)\langle T^{x_kx_k}\rangle,\\
\delta_Y [\sqrt{\gamma}\langle T^{tt}\rangle] &\approx \Big[ (1+\alpha \frac{i}{d})^{d+2-d}(1-\alpha \frac{d-1}{d})^2 -1\Big]\langle T^{tt}\rangle.
\eea
This gives the density in the boundary symplectic pairing
\bea
\label{eq:inftfdtderivativesympl}
\delta_Y [\sqrt{\gamma}\langle T^{ab} \rangle] \delta_t \gamma_{ab}-&\delta_ t [\sqrt{\gamma}\langle T^{ab} \rangle] \delta_Y \gamma_{ab} =\\ &=-2 \alpha f'(t) \frac{(  d-1 - i) \langle T^{tt}\rangle - (1 + i) (d-1) \langle T^{x_k x_k}\rangle}{d} \\
&=- 2 \alpha f'(t)\langle T^{tt} \rangle,
\eea
where in the last line we have used again that the background stress tensor is traceless and isotropic $(d-1)\langle T^{x_k x_k}\rangle=\langle T^{tt} \rangle$. 
Integrating this on half of the time fold yields the symplectic pairing\footnote{The integral measure is $dt_E=idt$.}
\beq
\Omega(\delta_Y,\delta_{\cal T}) =- i\lim_{{\cal T} \rightarrow \infty} V_{\rm bdy} \int_{-i\beta/4-{\cal T}}^{i\beta/4+{\cal T}} d(it) 2 \alpha f'(t)\langle T^{t t} \rangle = 2\alpha M [f(\infty)-f(-\infty)],
\eeq
which agrees with $2(d-1)\alpha$ times the volume derivative \eqref{eq:Tderivative}. 

As a final remark, note that the effects of the imaginary shift in the York time $\kappa$ in \eqref{eq:infdiffeo} cancel in both symplectic pairings \eqref{eq:betaderivativesympl} and \eqref{eq:inftfdtderivativesympl}. This is in line with the intuition that a shift in York time acts trivially on the boundary time slice and therefore it cannot produce cutoff independent contributions.

\subsection{Marginal deformations and the information metric}
\label{sec:tfdinfmetric}

Let us briefly return to the information metric \eqref{eq:infmetric} and its bulk expression \eqref{eq:infmetricbulk} in the context of the time evolved thermofield double state.
In particular, \cite{MIyaji:2015mia} calculated the time dependence of $G_{\lambda \lambda}$ in this state for $d=2$. They found that this time dependence is similar but not exactly the same as that of the extremal volume in the bulk.  The purpose of this section is to explain, using \eqref{eq:infmetricbulk}, why $G_{\lambda \lambda}$ has the same late time growth as the extremal volume. In addition, in appendix \ref{app:infmetricTFD} we reproduce the exact time dependence of $G_{\lambda \lambda}$ obtained in \cite{MIyaji:2015mia} by a bulk calculation using \eqref{eq:infmetricbulk}.

 In $d=2$, in global coordinates, the exterior is described by \cite{Hartman:2013qma}
\begin{equation}
\label{eq:2dtfdmetric}
ds^2=d\rho^2-\sinh^2 \rho dt^2+\cosh^2 \rho dx^2
\end{equation}
Instead of the sign deformation $\delta_s \lambda$ of \eqref{eq:infmetricdeformations}, we are going to work with an analytically continued deformation $\delta_h \lambda^-=1$, $\delta_h \lambda^+ =0$. It is clear that this also gives the information metric \eqref{eq:infmetric} when applied together with $\delta_c \lambda$ in the symplectic form \eqref{eq:qftkahler}. This change is merely to ease the formulas below, but one must keep in mind that this leads to complex Lorentzian solutions.
We can write the corresponding scalar profile using the massless bulk to boundary propagator in this background\footnote{The source deformation $\delta_h$ and the solution below was considered before in \cite{Trivella:2016brw}, but there the information metric was extracted directly from the on-shell Euclidean bulk action, rather than a symplectic flux on a Lorentzian Cauchy slice.}
\begin{equation}
\delta_h \langle \phi(\rho,t_E)\rangle_{TFD,\beta=2\pi,{\cal T}}=  \int_{-\frac{\pi}{2}-i {\cal T}}^{\frac{\pi}{2}+i {\cal T}} dt_E' \int dx \frac{A}{4 (\sinh \rho \cos (t_E-t_E')-\cosh \rho \cosh x)^2},
\label{mode} \end{equation}
where $A$ is some normalization constant ensuring that the propagator approaches a delta function near the boundary.
In appendix \ref{app:infmetricTFD}, we show explicitly how one gets the same time dependence as \cite{MIyaji:2015mia} from this bulk field. For general times, the analysis is not very illuminating, but for late times, something nice happens. At late times, the maximal volume surface is $\rho=i \frac{\pi}{4}$. In order for \eqref{eq:infmetricbulk} to give the volume as we approach this surface, we need $\delta_h \pi \approx i\partial_\rho \delta_h \phi$ to become a constant. In order for this to happen, we need the solution \eqref{mode} not to have $t$ dependence. So, this solution  has to satisfy $\partial_{\rho} (\sinh{\rho} \cosh{\rho} \partial_{\rho} \phi)=0 \rightarrow \phi=a_1+a_2 \log \tanh \rho$. Normally, the second solution cannot be generated in Euclidean signature, because it is singular near the horizon. So, naively it would seem that one cannot get this constant momenta mode at the extremal slice. However, if we take the large ${\cal T}$ limit of \eqref{mode}, we can perform the integral and we actually obtain the desired bulk mode: 
\bea
\delta_h \langle \phi(\rho,t)\rangle_{TFD,\beta=2\pi,{\cal T} \gg \beta } &\approx -i A (\log \tanh \rho- i \frac{\pi}{2}) , \\
& \Downarrow \\
\delta_h \pi(\rho=i \pi/4, t \ll {\cal T}) &\approx A.
 \eea
When we evaluate the solution at late Lorentzian times,  $t > {\cal T}$, we get again a constant and thus it does not contribute to the momenta. In this way, we recover from \eqref{eq:infmetricbulk} the linear growth of the volume with ${\cal T}$. This explains why the late time growth is the same as for the extremal volume. 

Now that we understand how the late time growth emerges in this model, we can analyse this approach from the boundary point of view in general dimensions. We have to evaluate 
\bea
\partial_{{\cal T}} \Omega(\delta_s \lambda, \delta_c \lambda) &=2 i  V_{bdy}  \int d^{d-1}x \int^{\infty}_{-\infty} dt G(t,x)\\ &=i 2  V_{bdy} G_{\Delta=d}(\omega=0,k=0)
\eea
where we have used $\partial_{\cal T}$ to get rid of one of the time integrals, in the other we have taken the large time limit and we got a $V_{bdy}$ for the integral over the sum of the positions. 
So, for large times, this symplectic form is proportional to the zero frequency, zero momenta limit of the marginal field two point function. In holographic theories, this is related with viscosity which is in turn related with entropy (energy) \cite{Policastro:2001y}. For holographic theories, this goes like $\frac{d}{2(d-1)} M$. Therefore, this quantity also displays the linear growth at late times. The factors of $d$ in front give a slightly different coefficient from the volume \eqref{eq:TFDvolume} in higher dimensions, it only gives the right answer in $d=2$.

\section{Complexity=volume?}
\label{sec:complexity}
\subsection{A new field theoretical definition of complexity}

In this section, we would like to provide a possible interpretation of our results in light of the complexity=volume conjecture of \cite{Susskind:2014rva,Stanford:2014jda}. Given that we work in the space of sources, it seems natural to consider distances in this space. That is, we want to consider the weighted distance in the space of sources, whose metric is the  K\"ahler metric coming from the K\"ahler potential \eqref{eq:kahlerpotqft}. Schematically,
\begin{equation}
\int_{s_i}^{s_f} ds F[g_{a b} \dot{\lambda }^a \dot{\lambda}^b],
\end{equation}
where we imagine $a,b$ to incorporate all variables that we need to sum and integrate over, $\lambda^a$ are coordinates on the complexified source space, and $g_{ab}$ is the symmetric variation of the partition function \eqref{eq:kahlerpotqft}, $g_{ab}=(\delta_a^+ \delta_b^-+\delta_b^+ \delta_a^-)\log Z[\lambda]$.
The most natural functionals are the kinetic energy $F[y]=y$ and the geodesic distance $F[y]=\sqrt{y}$.\footnote{We remind the reader that the extremal curves to this functional are the same geodesics for any choice of $F$, with the caveat that for $F[y]\neq \sqrt{y}$ we break reparametrization invariance, so we get the geodesics in a specific parametrization.} This is a weighted distance in source space between a reference state with sources $\lambda(s_i)= \lambda_i$ and our final desired state with sources $\lambda(s_f)= \lambda_f$. In order to evaluate the distance, we want to follow the minimal path.

How can we determine the right functional $F$ to use? It was suggested in \cite{Brown:2017jil}, in a slightly different context, that the kinetic energy $F[y]=y$ is the right functional because it is additive. If we have two decoupled CFT's with their respective sources, the partition function will be the product of the individual partition functions and thus the K\"ahler metric will be the sum of the metrics. We expect that a good notion of complexity should be additive in the sense that we should add up the complexities of tensor product states. This discussion also applies to the extremal volume in the bulk: if we have two disconnected AdS universes, the volumes add up. Furthermore, the complexity should scale like the number of local degrees of freedom, namely it should scale linearly with the spatial volume and $N^2$. These arguments pick out the kinetic energy as it is linear in the K\"ahler metric and we thus propose to define complexity of the Euclidean path integral states \eqref{eq:pathintstates} as
\begin{equation}
\label{eq:ourcomplexity}
{\cal C}(s_i,s_f)=\int_{s_i}^{s_f} ds g_{a b} \dot{\lambda}^a \dot{\lambda}^b.
\end{equation}

Now how does the volume of an extremal slice fit in this discussion? We can use the complex structure \eqref{eq:J} to rewrite our formula \eqref{eq:volumefromsympl} for the variation of the volume as 
\begin{equation}
\delta V \propto \Omega(\delta_Y  \lambda,\delta  \lambda)=g_{a b} J[{\delta}_Y { \lambda}]^a \delta { \lambda}^b \label{Cvol}
\end{equation}
We can obtain a similar expression from \eqref{eq:ourcomplexity}, by taking a variation with respect to the endpoint coordinate. If we do this for an on-shell trajectory, we only get a boundary term
\begin{equation}
\delta_{{ \lambda}_f}{\cal C}= \dot{\lambda}^a|_{\lambda_f} g_{a b} \delta { \lambda}_f^b. \label{Cyork}
\end{equation}
Therefore, it is natural to conjecture that $J[\delta_Y \lambda]=\dot{\lambda}^a|_{\lambda_f}$, that is, the image of the new York deformation \eqref{eq:weyl} under the complex structure $J$ should be identified with the tangent vector to the minimal trajectory in source space between $\lambda_i$ and $\lambda_f$.\footnote{Note that, if we had chosen another $F[y]$, the volume deformation would be non-linear in $\dot{\lambda}$ and the normalization would be different. We do not physically expect any of these two properties, so the kinetic energy seems to be singled out by demanding that the change in the volume just corresponds to a linear variation of the sources. } This would then imply that $\delta {\cal C} =\delta V$. Of course, this identification is highly speculative, and most of the remainder of this paper will be about gathering some evidence for it.

Before doing so, we need to complete the definition of \eqref{eq:ourcomplexity} by specifying the reference state $ \lambda_i$. Since $\cal C$ is positive and zero if $ \lambda_f=\lambda_i$, it clearly satisfies that $\delta_{\lambda_f}{\cal C}|_{ \lambda_f= \lambda_i}=0$, that is, variations around the reference state vanish. We have shown in section \ref{sec:vacuumdeformation} that around the vacuum, there are only divergent contributions to the variation of the volume, moreover that there is a boundary regulator for which this variation, as defined from the symplectic form, is zero.\footnote{This argument was for variations of the volume induced by a change of the boundary metric, which sources the stress tensor. If we turn on a source for another operator $O$, the variation around the vacuum trivially vanishes since the two point function $\braket{TO}$ is zero in the vacuum.} Therefore, unlike in most approaches to complexity in quantum field theory, we pick our reference state to be the state where all sources are turned off (namely $ \lambda_i=0$) which for states on the sphere will be the vacuum state.

\subsection{Relation to other approaches}

Here we briefly summarize how our approach relates to the extensive literature on defining complexity of states in a quantum field theory. Most work can be divided up along two major axes. The first is whether the prescription to count gates is independent of the initial state (this is usually called Nielsen's approach \cite{nielsen2005geometric}, a sample of works is \cite{Jefferson:2017sdb,Magan:2018nmu,Caputa:2018kdj,Chapman:2018hou}), or not (this is often based on the Fubini-Study metric, see e.g. \cite{Chapman:2017rqy}, and \cite{Ali:2018fcz} for a comparison between these two categories). Clearly, our prescription belongs to the second category, as it is also based on the Fubini-Study metric. A key difference from most of these works is that, as we will see, we can perform calculations without restricting to free fields, see however \cite{Magan:2018nmu} for a general approach based on the action of symmetry groups, and \cite{Caputa:2018kdj} for an application of this approach to Virasoro coherent states. 
The other axis is along whether we count only unitary gates (this is what \cite{Jefferson:2017sdb,Chapman:2017rqy,Magan:2018nmu,Caputa:2018kdj,Chapman:2018hou} follow) or introduce some notion of counting non-unitary gates. Works in the second category are usually based on some notion of counting gates in the preparation of the state with a Euclidean path integral \cite{Caputa:2017urj,Caputa:2017yrh,Czech:2017ryf,Bhattacharyya:2018wym,Takayanagi:2018pml}, which is in spirit fits very well with what we are doing, but unlike our approach, these works are not based on distance functionals. Another important difference from the path integral optimization story of \cite{Caputa:2017urj} is that since we build our geometry using the Fubini-Study metric, our complexity functional is completely blind to the normalization of states, while \cite{Caputa:2017urj} defines the optimal circuit by minimizing the normalization. 

\subsection{Ba\~nados geometries}

A natural testing ground of the above conjecture is conformal deformations of the vacuum state in $d=2$ CFTs, partially because this provides a nontrivial setup where the answer is fixed by symmetry, and partially because there are available results in the literature for the volume of the extremal slice.

We begin by computing the complexity \eqref{eq:ourcomplexity} of a state created from the vacuum by applying a small conformal transformation\footnote{Note that we use the ``bad" convention $x^{\pm} = t \pm x$.}
\bea
x^{\pm} = \tilde x^{\pm}+\sigma g_{\pm}(\tilde x^\pm).
\eea
We are going to work to leading nonvanishing (i.e. quadratic) order in $\sigma$.
These states can be explicitly written as
\beq
\label{eq:conftrafstate}
| g_+ g_-\rangle = U_{g_+} U_{g_-}|0\rangle,
\eeq
where we act with the unitary Virasoro representation
\bea
\label{eq:virasoro}
 U_{g_+} U_{g_-} &\approx \exp \left( i \sigma  \int dx g_+(x) T_{++}(x) + g_-(-x)T_{--}(x) \right) \\
 &\equiv \exp \left( i  \int dx  t_f(x)T_{tt}(x) +  x_f(x) T_{tx}(x) \right).
\eea
In the above formula, we introduced the parametrization
\bea
\label{eq:banadosdeltax}
 t_f(x)&=\frac{\sigma}{2}[g_+(x)+g_-(-x)], && \quad &&
x_f(x)&=\frac{\sigma}{2}[g_+(x)-g_-(-x)],
\eea
which will be useful later. $t_f(x)$ is essentially the shape of the $t=0$ surface after the coordinate transformation, while $x_f$ is the longitudinal shift along this surface. The K\"ahler potential \eqref{eq:kahlerpotqft} should be obtained from the norm
\beq
\mathcal{K}=\log \langle g_+ g_-|g_+ g_-\rangle,
\eeq
where we need to complexify the Euclidean sources creating the state, so that this norm is not trivial. It is not immediately obvious how to do this, since the $t=0$ surface is given by a non-holomorphic constraint.
The most naive thing to do would be to just complexify both $g_+$ and $g_-$ independently. But this is too much, since they determine the entire background geometry, so this would give a complex background metric without the required $Z_2+C$ symmetry in \eqref{eq:lambdatilde}. Instead, we should examine the background metric giving rise to this transformation
\bea
ds^2
&=[1+\sigma g_+'(x^+)][1+\sigma g_-'(x^-)]dx^+ dx^- \\
&\approx (1+\sigma[ g_+'(x^+)+g_-'(x^-)])dx^+ dx^-.
\eea
Going over to Euclidean and enforcing symmetry under $Z_2+C$, we get that 
\beq
 g_+'(x)+g_-'(-x)=[ g_+'(x)+g_-'(-x)]^*
\quad \Rightarrow \quad
 x'_f(x)=[ x'_f(x)]^*.
\eeq
Therefore, the K\"ahler potential should be obtained by complexifying $t_f$ but leaving $x_f$ real.\footnote{There could be a complex constant mode in $x_f$ but this is just a translation and it annihilates the vacuum.} This leads, to leading nonvanishing order, to the K\"ahler potential
\bea
\mathcal{K}&=\log \langle g_+,g_-|g_+,g_-\rangle_{x_f\text{ real, }t_f \text{ complexified }} 
\\ &\approx -\frac{1}{2}\int dx dy \frac{c}{(x-y)^4}[t_f(x)-t_f^*(x)][t_f(y)-t_f^*(y)],
\eea
where the kernel comes from the vacuum two point function of the stress tensor.
The geodesic connecting the vacuum to $t_f(x)$ in the corresponding K\"ahler metric is just a straight line
\beq
t(s,x)=t_f(x) \frac{s}{s_f} , \label{straight}
\eeq
Therefore, the complexity of these states read as
\bea
\mathcal{C}&=\int_0^{s_f} ds \int dx dy \frac{c}{(x-y)^4}\partial_s t(s,x)\partial_s t^*(s,y)\\ & = \frac{1}{s_f}\int dx dy \frac{c}{(x-y)^4}t_f(x)t_f^*(y).
\eea

The next thing to do is to compare to the volume of the maximal Cauchy slice in the AdS$_3$ geometry dual to the state \eqref{eq:conftrafstate}. This state is dual to the Ba\~nados geometries of \cite{Banados:1998gg}, and luckily the leading order finite change in the volume compared to vacuum AdS was calculated recently in \cite{Flory:2018akz}. As written in this reference, the result is
\beq
V^{(2)}=\sigma^2 \pi^3 \int_{-\infty}^{\infty} d\xi |\xi|^3 [\hat g_+(\xi)+\hat g_-(-\xi)][\hat g_-(\xi)+\hat g_+(-\xi)] \,,
\eeq
where we have set their anchoring time $t_0$ to zero without the loss of generality. The $\hat g$ functions are related to the $g$s by Fourier transformation 
\beq
\hat g_{\pm}(\xi)=\frac{1}{2\pi} \int dx e^{2\pi i x\xi} g_{\pm}(x).
\eeq
We can easily rewrite this formula in real space. Neglecting prefactors we obtain
\bea
V^{(2)}&\propto \sigma^2 \int dx dy \frac{1}{(x-y)^4}[g_+(x)+g_-(-x)][g_+(y)+g_-(-y)]\\
&=4\int dx dy \frac{1}{(x-y)^4}  t_f(x) t_f(y).
\eea
Comparing with the field theory calculation, we see that\footnote{We note that there is another notion of complexity for states that are conformal transformations of the vacuum in a $d=2$ CFT based on the Kirillov-Konstant action on the Virasoro coadjoint orbit \cite{Caputa:2018kdj}. This action is local instead of bilocal in the large $c$ limit, therefore it seems unlikely to us that it could reproduce the above change in the volume. }
\beq
\mathcal{C} \propto \frac{1}{s_f} V^{(2)} , \quad \quad \delta_{t_f} \mathcal{C} \propto \frac{1}{s_f} \delta V^{(2)}.
\eeq

\subsection{Mini-superspace approximation for the TFD}

As a simple toy model, we can consider the family of states defined by thermofield doubles at different temperatures and times. That is we can define 
\beq
|\tau \rangle=e^{-\tau H} |TFD(\beta)\rangle,
\eeq
where $\tau$ is a complex parameter and we use $\beta$ as a reference temperature. We complexify the parameter $\tau$ (which can be though of as a source for a diffeo), which gives rise to the K\"ahler potential $\mathcal{K}=\log \langle \tau| \tau \rangle = \log Z(\beta+2 \tau+2 \tau^*) -\log Z(\beta)$. We therefore have a two dimensional space defined by $\tau,\tau^*$, with the following metric:
\bea
ds^2&=\partial_{\beta}^2 \log Z(\beta+2 \tau+2 \tau^*) d\tau d\tau^* \\ &=\frac{f d}{(\beta+2 \tau+2\tau^*)^{d+1}} d\tau d\tau^*.  \label{minisuper}
\eea
Here, we have used that the thermal partition function on the plane is fixed by dimensional analysis, up to a coefficient $f$ proportional to the spatial volume 

Let us first focus on the Lorentzian time evolution submanifold with $\tau+\tau^*=0$.
Given that the kinetic energy is not reparametrization invariant, following \cite{Susskind:2018pmk}, it seems natural to choose the parametrization in terms of ``Rindler time" $\tau(s)= \beta s$. This leads to the distance between $\tau(0)=0$, $\tau(s_f)={\cal T}$ 
\bea
\int ds g_{a b} \dot{x}^a \dot{x}^b=M {\cal T} && \quad && M=\langle H \rangle_{\beta}=f \beta^{-d},
\eea
which gives the right growth, so it is a further justification for using the kinetic energy in \eqref{eq:ourcomplexity}.\footnote{The geodesic distance does not have the right growth: $\int ds \sqrt{g_{a b} \dot{x}^a \dot{x}^b}=\sqrt{d M /\beta} T $.} However, notice that using the Rindler time as the parameter of the geodesic leads to an $s_f$ that depends on the final state, and therefore it is in tension with the argument that we gave to connect the complexity \eqref{eq:ourcomplexity} to the extremal volume. The previous example of the Ba\~nados geometries does not suffer from this problem.

\subsubsection*{Geodesics in mini-superspace}

We can actually get a non-trivial geodesic using the metric \eqref{minisuper}, with quadratic early time growth and non-trivial intermediate time dynamics. The exact time dependence we obtain in this section differs from that of the actual volume, but it is qualitatively close. Also, because we are looking for geodesics in a positive definite metric, we expect the actual value of $\cal C$ to be less than what we will obtain here. The calculation we present in this section is also a nice example of a field theoretic calculation of complexity that does not rely on using free fields. 

We set $\beta/4+\tau = y+i x$, $\beta/4+\tau^* = y-i x$ and we look for geodesics starting at $y=\beta/4,x=0$ and ending at $y=\beta/4,x={\cal T}$, that is, TFD at time zero and TFD at time ${\cal T}$. Our distance functional \eqref{minisuper} becomes
\beq
\mathcal{C} \propto \int ds \frac{\dot{x}^2+\dot{y}^2}{y^{d+1}}.
\eeq
The action is $s$ independent so the Hamiltonian is conserved, moreover it is $x$ independent so the canonical momentum of $x$ is conserved. This gives the EOMs
\bea
\label{eq:quadreom}
\dot {x} = \frac{1}{2} p_x y^{1+d}, && \dot{y}=\frac{1}{2}\sqrt{4 k y^{1+d}-p_x^2 y^{2+2d}},
\eea
where $p_x$ and $k$ are constants of integration. The meaning of $k$ is the value of the Hamiltonian of the system, but it is also just the norm of the tangent vector to the curve (i.e. the value of the Lagrangian): 

\beq
k=\frac{\dot{x}^2+\dot{y}^2}{y^{d+1}}.
\eeq
This implies that the rate of change of $\mathcal{C}$ with respect to the circuit time $s_f$ is constant
\beq
\frac{d}{ds_f} \mathcal{C}= k.
\eeq
 We now fix the constant $k=\beta^{1-d}$ based on dimensional grounds. This is the step where we essentially select Rindler time as the parameter of the geodesic. The important point is that the rate of change is less trivial with respect to physical time, which is defined as $x(s_f)={\cal T}$. 

\begin{figure}[h!]
\centering
\includegraphics[width=0.55\textwidth]{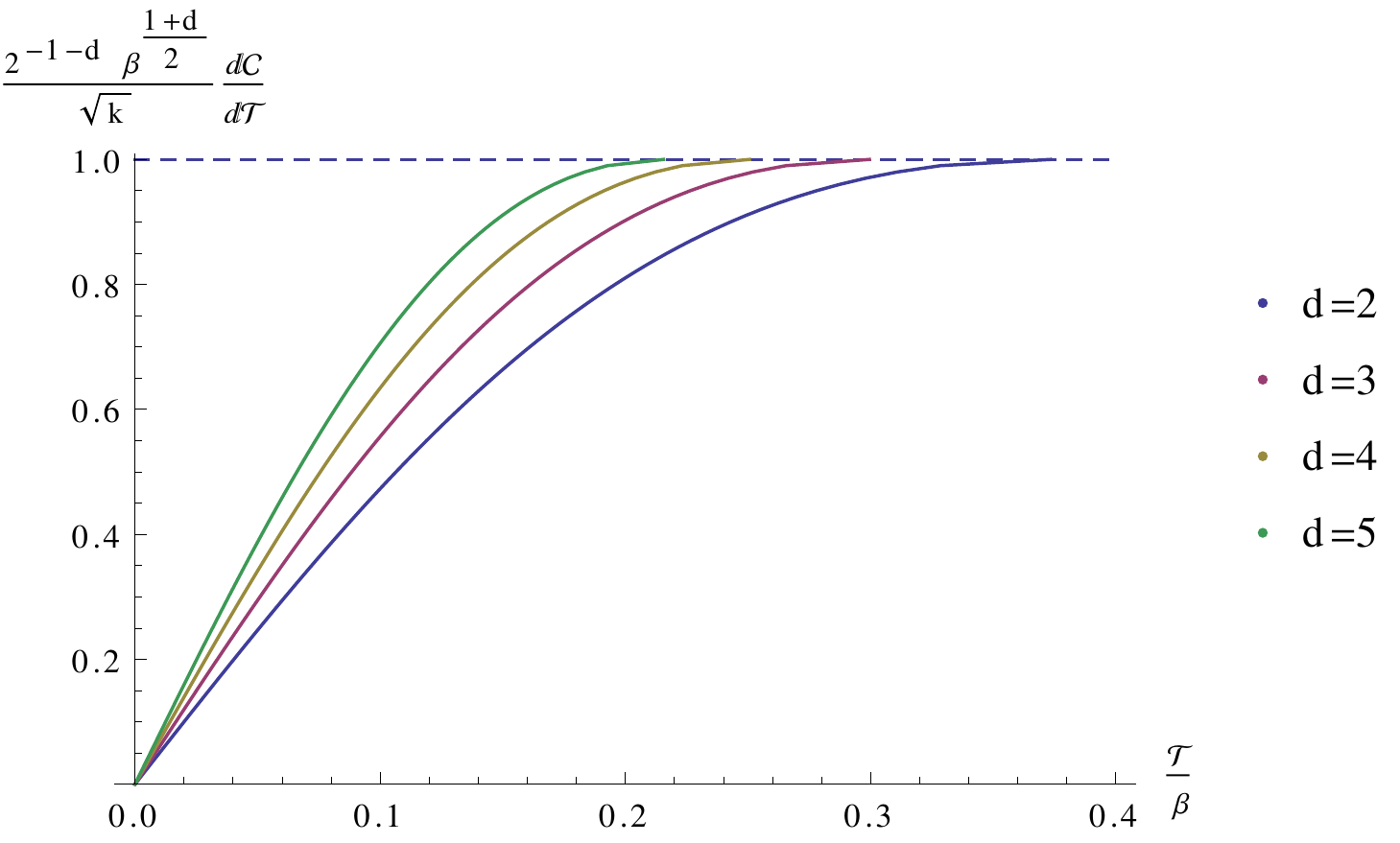} 
\caption{Plot of \eqref{eq:px2} for $d=2,...,5$, larger $d$ hugs the infinite time line more.}
\label{fig:4}
\end{figure}

Now let us discuss the solutions of \eqref{eq:quadreom}. We will show that there are two competing solutions and they exchange dominance at some time ${\cal T}\sim \beta$. The generic solution will have a turning point where $\dot y=0$, i.e.
\beq
y=y_*=\left( \frac{4 k}{p_x^2}\right)^{\frac{1}{d+1}}.
\eeq
By symmetry, the turning point happens at $s=s_f/2$, where the imaginary part must be $x(s_f/2)={\cal T}/2$, which means that $s_f/2=\int^{{\cal T}/2}dx (\dot x)^{-1}$ and therefore $\frac{d}{d{\cal T}}s_f=(\dot x)^{-1}_{\text{turning point}}=2(p_x y_*^{1+d})^{-1}$. Since we fixed a $k$ independent of ${\cal T}$,
the extremal value of the functional will satisfy
\beq
\frac{d}{d{\cal T}} \mathcal{C}=2\frac{ k}{p_x y_*^{1+d}}=\frac{1}{2}p_x,
\eeq
so the rate of change of the extremal complexity with respect to the physical time is the other conserved charge $p_x$.
Now the simplest solution is setting
\bea
\dot{y}=0, && y=y_*,
\eea
along the total curve. Since we required $y(0)=y(s_f)=\beta/4$ we must have
\beq
\label{eq:px1}
y_*=\beta/4 \;\; \rightarrow  \;\; p_x=2^{2+d}\frac{ \sqrt{k}}{\beta^{\frac{1+d}{2}}}
\eeq
The condition $x(s_f)={\cal T}$ just fixes $s_f$ in terms of $k$ and $\beta$, while we see that our choice $k\propto \beta^{1-d}$ leads to
\beq
p_x \sim \beta^{-d} \sim M,
\eeq
which is the right linear growth. So we see that the previously considered trajectory, where $\tau$ has no real part, is a good geodesic for this two dimensional space as well. However, there is another solution, where $y=y_*$ is a genuine turning point. In this case, $p_x$ is not fixed by \eqref{eq:px1}, but instead by requiring that
\bea
\label{eq:px2}
{\cal T}/2 &=\int_{\beta/4}^{y_*}dy \frac{dx}{dy} \\
&=\int_{\beta/4}^{y_*} dy \frac{p_x y^{1+d}}{\sqrt{4k y^{1+d}-p_x^2 y^{2+2d}}} \\
&=\frac{p_x \beta^{\frac{3+d}{2}}}{2^{3+d}(3+d)\sqrt{k}} {}_2 F_1\left(\frac{1}{2},\frac{3+d}{2+2d},\frac{5+3d}{2+2d};\frac{4^{-2-d}p_x^2 \beta^{1+d}}{k} \right).
\eea
We plot the associated result for complexity in Fig. \ref{fig:4}. For small times, this solution is better than \eqref{eq:px1}. This can be seen by expanding for small $p_x$:
\beq
\label{eq:tsquare}
{\cal T}/2 = \frac{2^{-3-d}\beta^{\frac{3+d}{2}} p_x}{(3+d)\sqrt{k}} + \cdots \;\; \rightarrow \;\; p_x \sim \beta^{-1-d}{\cal T},
\eeq
where we have inserted again $k\sim \beta^{1-d}$. Note that this implies $\mathcal{C} \sim \beta^{-1-d}{\cal T}^2$ at early times, therefore it is smaller than the value of the functional on the solution \eqref{eq:px1}, which has $\mathcal{C} \sim \beta^{-d}{\cal T}$. 

The r.h.s. of \eqref{eq:px2} grows monotonically in $p_x$ until it becomes imaginary precisely when $p_x$ reaches its value fixed by the other solution $\eqref{eq:px1}$, or equivalently, the argument of ${}_2 F_1$ becomes one. So the two solutions exchange dominance precisely when \eqref{eq:px2} stops to exist. This is a typical second order phase transition behaviour. Evaluating the r.h.s. at this point gives the critical time where the solution \eqref{eq:px2} is replaced by the solution \eqref{eq:px1}
\beq
\label{eq:criticalt}
{\cal T}_{\rm crit}=\frac{\sqrt{\pi} \beta \Gamma(\frac{3}{2}+\frac{1}{1+d})}{(3+d)\Gamma(1+\frac{1}{1+d})}.
\eeq

How does this time dependence compare to the time dependence of the extremal volume in the eternal wormhole? Since our model does not fix the overall coefficient of $\mathcal{C}$, we can only meaningfully compare e.g. the ratio of the coefficient of the early time ${\cal T}^2$ and the late time ${\cal T}$ growth. In other words, we choose a coefficient such that the late time growth matches and then compare early times. For our toy model the ratio comes from comparing \eqref{eq:px1} and \eqref{eq:tsquare}:
\beq
\frac{\frac{d}{d{\cal T}}\mathcal{C}({\cal T}\ll 1)}{\frac{d}{d{\cal T}}\mathcal{C}({\cal T} \rightarrow \infty)}=\frac{3+d}{\beta}{\cal T}.
\eeq
For the actual volume, we can use the results of \cite{Kim:2017qrq} to get
\beq
\frac{\frac{d}{d{\cal T}}V({\cal T}\ll 1)}{\frac{d}{d{\cal T}}V({\cal T} \rightarrow \infty)}=\frac{2\sqrt{\pi} d \Gamma(\frac{1}{2}+\frac{1}{d})}{\beta \Gamma(\frac{1}{d})}{\cal T}.
\eeq
The coefficient is larger in the toy complexity for any $d\geq 1$. This is reassuring, because we expect the true geodesic energy to be smaller than the one obtained by the mini superspace approximation.

\section{Conclusions}
\label{sec:disc}

In this paper, we have first discussed how to interpret the new York deformation which generates the volume by understanding its interpretation in GR and studying some explicit examples.  Of course, one of the main challenges is to understand better the boundary interpretation of $\delta_Y \gamma$: while it is uniquely determined from the bulk and we have computed it explicitly in several examples, it is not clear to us what the best way is to motivate this object purely in terms of the boundary, other than the complexity motivated conjecture \eqref{Cvol}=\eqref{Cyork}. It might be that there is some precise way to think about the boundary dual of time evolving the gravitational variables, while keeping the surface fixed. 
 
 In the context of the complexity=volume conjecture, our setup leads to a natural proposal for complexity in terms of the minimal kinetic energy between two states, measured in the space of Euclidean sources. We have checked some examples  such as a minisuperspace thermofield-double setup and perturbative Ba\~nados geometries. We found qualitative agreement, although in the TFD case, we needed a state dependent parametrization. Note that it is very hard to carefully check \eqref{Cvol}=\eqref{Cyork} since it would require studying arbitrary source deformations, and finding the minimal path in an infinite dimensional space. For practical reasons, we are therefore quickly forced into mini-superspace examples. Nevertheless, we believe the agreements we have found are non-trivial and it would be interesting to explore this notion of complexity further and hopefully prove \eqref{Cvol}=\eqref{Cyork}. 
 
 Below we expand on some of the open questions and possible future work.

\subsection*{A brane description of the volume and analogies with entanglement}

From our experience with entanglement entropy, where the area of a codimension-2 surface can be computed in terms of adding a fictitous probe brane to the geometry \cite{Lewkowycz:2013nqa}, it would be nice if there was an analogous probe brane construction which gave us the volume.

From our current description in terms of the new York deformation $\delta_Y \gamma$, it seems difficult to make this connection, since we do not have an explicit state independent formula for this deformation. However, in the case of the vacuum, we have seen that the boundary deformation is a sign function, which is singular in the boundary, but is regulated in the bulk as it is common in the entanglement entropy context (see for example \cite{Camps:2016gfs}). In a similar way to \cite{Lewkowycz:2013nqa},  one might be able to understand if there is some brane picture by using the $Z_2+C$ symmetry to orbifold our system: this will give a \textit{smooth} boundary geometry with a physical boundary at $t=0$, which will seem to require a physical brane in the bulk which ``absorbs" the remaining momenta at $\Sigma$. From this point of view, the new York deformation is equivalent to a small change of the brane tension. Maybe this can be made more precise using the ideas of \cite{Takayanagi:2011zk}.

In some sense, the vacuum discussion of section \ref{sec:vacuumdeformation} is the volume equivalent of the Casini-Huerta-Myers map \cite{Casini:2011kv}. The new York transformation is simple in WdW coordinates, whose natural boundary geometry is the hyperboloid, which can be obtained by a conformal transformation of the boundary, where we send the $t=0$ surface to infinity. While the same words apply to \cite{Casini:2011kv}, in our situation, we only have access to $\delta V$, which in this case was trivial. We still expect that the vacuum picture will be helpful to understand $\delta_Y \gamma$ more generally: for example, it probably characterizes the near $t = 0 $ behaviour of other states with spherical topology, this near defect behaviour was all we needed to define the entanglement entropy.

\subsection*{Divergences and ambiguities}

We have seen that around the vacuum, the new York transformation agrees with a translation in York time, which is a diffeomorphism. Therefore, the variation of the volume is a boundary term. Moreover, since York time is an internal time to the WdW patch, it does not move the boundary time slice, and therefore this boundary term is only nontrivial because we need to cut off the geometry. In terms of the CFT, it means that it is only there due to the presence of the UV cutoff.
However, we were not able to relate the boundary and bulk regulators in any natural way. It is well known that Wald's symplectic form is ambiguous up to the addition of a boundary symplectic form, which is known as the JKM ambiguity \cite{Jacobson:1993vj}. We expect that the choice of this boundary term maps to the ambiguity in the choice of the regulator. Note that one could imagine boundary terms that give finite contributions to the symplectic form, so we are not claiming that the finite piece in the volume is always universal. But the point is that they can only depend on a finite number of derivatives of the sources on the $t=0$ surface. Therefore, for sources that are switched off sufficiently quickly, these boundary terms do not contribute. On such a class of states, it might make sense to define a finite part of the volume which is universal. It would be interesting to understand this better.

\subsection*{Quantum corrections}

Our symplectic forms are quadratic in the field variations. This does not mean that we have to restrict to leading order in $N$ for the variations. If we deform the state by a source $\delta \lambda \sim O(1)$, the change in the expectation value will have an expansion in $1/N$: while it will still be given by the integral of the two point function over the boundary source, this two point function has a $1/N$ expansion. So, since in the boundary this object is well defined in QFT, the symplectic form is well defined to arbitrary order in the $1/N$ expansion: it is still going to be quadratic in $\delta \lambda,\delta \langle O \rangle$, but $\delta \langle O \rangle$ gets corrections.

In the bulk, the same story is true: the boundary symplectic form is equal to the bulk symplectic form, if we change the classical variations for their quantum expectation values: $\omega(\delta_1 \langle \phi \rangle, \delta_2 \langle \phi \rangle)$.  The ``tadpole" contributions that determine the one point functions can be thought as the equations of motion for the background field stemming from the quantum effective action. In this way, since the one point functions will also satisfy the semiclassical ``tadpole" equations of motion, we can move this symplectic form to the $t=0$ bulk slice and we have that the quantum generalization of this story is simply exchanging classical variations by the variations in the expectation values: $\delta \phi \rightarrow \delta \langle \phi \rangle$. 

In the context of the volume, if we only  new York transform the metric but we have matter fields,  we will get the volume plus a contribution from the scalar fields, as was described in section \ref{sec:scalar}. This contribution might resemble the quantum corrections to the entanglement entropy of \cite{Faulkner:2013ana}. However, a difference from entanglement entropy is that, if the matter fields have a semi-classical expectation value, this matter contribution will be the same order as the leading volume. Nevertheless, as discussed in section \ref{sec:scalar}, we can also get the volume alone by adding the appropriate deformation of the other sources. From this point of view, we expect that by considering the appropriate new York deformation, there will not be any quantum corrections to $\delta V$, instead, the respective $\delta g$ will be the ``tadpole" contribution of the volume, i.e. $\delta g=\delta \langle \hat{g} \rangle$. However, as discussed later, in the context of complexity, it might be more natural to only source the metric but not the scalar sources (which is what one does in entanglement). 

\subsection*{Higher derivatives}

Our discussion about symplectic forms did not assume any particular gravitational Lagrangian. So the equality between bulk and boundary symplectic forms stays the same. However, it is not clear what the natural higher derivative generalization of the volume is. To our knowledge, there is no analogue of a York construction that could be used to describe  the gauge invariant phase space of higher derivative gravity (and writing the symplectic form in phase space variables is subtle: at every point, one has to specify not only first but also higher derivatives of the metric). One could take the deformation \eqref{eq:weyl} at face value and define the volume from the symplectic form $\delta V_{\rm higher} =\Omega_{\rm higher}(\delta_Y g,\delta g)$, but it is not clear to us if this is the right thing to do. The proposal of \cite{Bueno:2016gnv} for the generalized volume appears to be equivalent to doing the latter. 

\subsection*{Towards complexity=volume}

We have seen that we can define a complexity like quantity from the Fubini-Study metric on the space of classical states, and made some observations, based on the equivalence of bulk and boundary symplectic forms, which suggest that such a complexity could compute the volume of the maximal slice. These arguments relied on the state independence of the final ``computer time", $s_f$. In the case of Ba\~nados geometries, where we have found an exact match, this condition is satisfied. However, for the mini superspace approximation of the thermofield double state, it seemed more natural to pick Rindler time as the parameter $s$, in which case $s_f={\cal T}/\beta$ is state dependent. Since the kinetic energy complexity breaks reparametrization invariance, its value depends on this choice of $s_f$ and it would be interesting to understand better what the right choice is.

We should note that we are far from sure that our version of the complexity=volume conjecture is correct. The matching in the case of Ba\~nados geometries seems nontrivial (recall the crucial role of $Z_2+C$ symmetry), still, the form of the result is simple enough that one could argue for it only on symmetry grounds. We expect that performing these calculations in higher dimensions and/or at higher orders would give valuable insights. We hope to return to these questions in the future.

It is also important to point out that the conjecture, as it is stated, can only hold for states that are obtained by sourcing the stress tensor. This can be seen by considering scalar condensates along the lines of section \ref{sec:scalar}. The leading order complexity of such states is forced to be proportional to $\lambda^- \lambda^+$, because the geodesics in source space in this case are straight lines. However, as shown in section \ref{sec:scalar}, the change in the volume is given by the WDW energy of the scalar which can contain $(\lambda^-)^2$ and $(\lambda^+)^2$ terms. It could be that in the presence of matter, the natural object related to complexity is a combination of the volume and the matter Hamiltonian, in a similar spirit to how quantum corrections to entanglement entropy work, as was discussed in the a few subsections above. This would give a possible way around the problem above.

\subsection*{Comparison with HKLL}

Since we have only given an implicit expression for $\delta_Y$ in terms of the bulk fields, one might be tempted to compare it with the expression that one obtains by writing the bulk $\delta h^i_i$ in terms of boundary fields using HKLL \cite{Hamilton:2006az}. This gives a boundary expression that is intrinsically more complex than our formula, for one, it necessarily involves an integral over the bulk surface that does not simplify and the resulting boundary interpretation will just be a complicated linear combination of the stress tensor one point function for all Lorentzian times. In fact, while for entanglement entropy one could also do HKLL for $\delta A$, this expression does not present properties which are desirable of $\delta A$, for example, it depends on the stress tensor one point function everywhere as opposed to depending only on fields in the respective subregion. For entanglement entropy, the natural boundary expression for the area is $\delta A=\delta \partial_n \log Z$, where $\partial_n$ is an infinitesimal conical deformation.
Our situation is similar, we have that $\delta V \sim \delta \delta_Y \log Z=\langle \delta_Y \Psi|\delta \Psi\rangle $,  properly antisymmetrized. This expression is quite different from what one gets from HKLL (among others we have explicitly use extremality) and it can be naturally written just in terms of one boundary integral. We expect that proving the complexity=volume conjecture of section \ref{sec:complexity} or understanding better how $\delta_Y \gamma$ is determined by the local behaviour close to $t=0$ beyond the vacuum state will shed more light on the interpretation of this boundary object. 

\subsection*{On the definition of the path integral states and the Euclidean boundary problem in the bulk}

In this work, we addressed local issues around known backgrounds of the gravitational phase space in AdS and its encoding in the CFT Hilbert space. It would be nice to gain some understanding about global questions and in this subsection we would like to point out how surprisingly little is known about this.

From a purely CFT point of view, one could raise issues with the formal definition of path integral states \eqref{eq:pathintstates}. For a Lagrangian CFT, we can in principle define them by introducing a regulator and computing the path integral. Standard RG arguments then suggest that the symplectic form and the K\"ahler metric are cutoff independent, given that the source profiles are switched off in a buffer zone near $t=0$ that is much larger than the cutoff scale (and of course that they change slowly compared to the cutoff scale). This is because these objects are defined from the connected two point function in the deformed theory. Since the legs of the two point function are on opposite sides of the $t=0$ surface, it probes distances that are strictly larger than the size of this buffer zone, and in a renormalizable QFT such correlators are expected to be cutoff independent.

In the bulk, the equality of bulk and boundary symplectic forms is a simple consequence of the Euclidean extrapolate dictionary and can therefore be applied whenever we know how to use the latter. However, there is very little known about the set of boundary conditions that give rise to smooth Euclidean solutions in the bulk, which is still an open problem in General Relativity. Nevertheless, as long as the boundary partition function is well defined, we expect that the bulk solutions will be smooth, although this might require turning on extra fields in the bulk, for a Lorentzian example of this see \cite{Crisford:2017gsb}.

There could also be a problem with sourcing irrelevant operators. In this case, one expects the fully backreacted solution to not be asymptotically AdS, since the irrelevant deformation deflects the RG flow from the fixed point.\footnote{For states where sources are switched off near the $t=0$ surface, both the surface and the Lorentzian continuation are AAdS. } In such a scenario for generic sources, it is not entirely clear how to impose boundary conditions and apply the extrapolate dictionary. One usually thinks about this question in perturbation theory, where it is fine, since in this case we can keep the FG gauge and correct the metric order by order. This can lead to more singular growth in the FG coordinate $z$ than $z^{-2}$, but we can always cut off the geometry such that these terms are small. In this sense, the asymptotics are close to being AdS, or nearly-AdS and this procedure is essentially the same as the one used to deal with the dilaton in AdS$_2$ \cite{Maldacena:2016upp,Jensen:2016pah,Engelsoy:2016xyb}. However, we can clearly come up with nice and smooth but non-perturbative AAdS initial data for heavy massive fields in the bulk (for example deforming the five-sphere on $AdS_5\times S^5$ which would lead to a whole tower of irrelevant operator deformations). This suggests that there should be an understanding of the Euclidean boundary problem in this case. It would be interesting to understand these questions better.

\section*{Acknowledgements}
We are happy to acknowledge discussions with Vijay Balsubramanian, Jan de Boer, Matt DeCross, Ted Jacobson, Arjun Karr, Zohar Komargodski, Rob Myers, Onkar Parrikar, Tadashi Takayanagi, Mark Van Raamsdonk and Erik Verlinde.
AL and GS acknowledges support from the Simons Foundation through the It from
Qubit collaboration. AL would also like to thank the Department of Physics and Astronomy
at the University of Pennsylvania for hospitality during the development of this work. GS was partially supported by FWO-Vlaanderen through project G044016N, 
and by Vrije Universiteit Brussel through the Strategic Research Program 
``High-Energy Physics". AB is supported by the NWO VENI  grant  680-47-464  /  4114.
\appendix
\section{ADM and WdW}
\subsection*{ADM equations in general dimensions}
\label{app:adm}

Let us recall the ADM decomposition of general relativity in arbitrary dimensions and with a cosmological constant, see e.g. \cite{Witek:2013ora}.
We decompose the $d+1$ dimensional metric as
\beq
\label{eq:ADMcoord}
ds^2=-(N^2-N^a N_a)dt^2+ 2 N_a dx^a dt+h_{ab}dx^a dx^b,
\eeq
where $N$ is the lapse and $N^a$ is the shift, $h_{ab}$ is the induced metric of the constant $t$ surfaces. The indices above run over $d$ values, and are raised/lowered with $h_{ab}$. The equations of motion in this decomposition boil down to $d+1$ constraint equations
\bea
\nabla^a K_{a b}-\nabla_b K^{a}_{a} &=0,\\
 K_{a b} K^{a b}-(K_a^a)^2-R_d+2\Lambda&=0,
 \eea
 where $K_{ab}=\nabla_a n_b$ is the extrinsic curvature of the constant $t$ surfaces.
In addition to the constraints, the evolution of the $h_{ab}$ and $K_{ab}$ is determined by the equations
\bea
\label{eq:ADMeom}
\partial_t h_{ab}&=-2 N K_{ab}+\mathcal{L}_{N^c}h_{ab} \\ 
\partial_t K_{ab} &= -\nabla_a \nabla_b N + N(R_{ab}-2{K_a}^cK_{cb}+K K_{ab})+\mathcal{L}_{N^c} K_{ab}-\frac{2\Lambda}{d-1}N h_{ab},
\eea
where $\mathcal{L}$ denotes the Lie derivative. In this work, we will be focusing on negative cosmological constant and set the AdS radius to unity, leading to $\Lambda = -\frac{1}{2}d(d-1)$.

\subsection*{General discussion of the $Z_2$ symmetric case}

We have seen that time translation along York time gives rise to the new York transformation \eqref{eq:weyl} in vacuum AdS. Here we explore a bit further what are the conditions on the spacetime for York flow to give rise to \eqref{eq:weyl}. We clearly need that the conformal metric does not change along the flow. In ADM coordinates \eqref{eq:ADMcoord} with zero shift, this gives
\begin{equation}
0=\partial_{t} \bar{h}_{a b}=K_{a b}-\# K
\end{equation}
 and thus for extremal surfaces (which we needed in the first place), this will only be zero if $K_{a b}=0$, i.e. in $Z_2$ symmetric states. In these states, the Hamiltonian constraint in the background then forces $R_{d-1}=-d(d-1)$ (and its first time derivative). 
In order for the ADM slicing \eqref{eq:ADMcoord} to coincide with a CMC slicing we need $\partial_t K=1$ and the last equation we need is $\partial_{t} \bar{\pi}_{ab}=0 \rightarrow \partial_{t} K_{a b}=\frac{h_{a b}}{d}$. Using the EOM \eqref{eq:ADMeom} for $K_{a b}$, these equations become:
 \begin{equation}
( -\nabla^2 N+N d)= 1,~~  -\nabla_{a} \nabla_b N+N (R_{a b}+d h_{a b})= \frac{h_{a b}}{d}, 
 \end{equation}
 We are not going to attempt an exploration of solutions to these equations, but we just note that if we impose that $N=d^{-1}$ is a constant, then $R_{a b}=-(d-1)h_{a b}$. 
 These are basically WdW geometries \eqref{eq:WDWADS}, which usually correspond to the vacuum. 

\subsection*{WdW to Poincare coordinates}
\label{app:wdw}
In this section, we discuss the Wheeler-deWitt coordinates on AdS and their relation to the Poincare patch. The Wheeler-deWitt slicing of AdS$_{d+1}$ is given by the metric\footnote{In AdS$_3$, one may replace the hyperbola by $\mathbb{H}_2/\Gamma$ , a quotient with respect to a discrete subgroup of the hyperbolic symmetries. This corresponds to multi-boundary black holes or pure states with topology behind the horizon \cite{Maloney:2015ina}.}
\be \label{WdWmetric}
ds^2=-d\tau^2 + \cos^2\tau \  \frac{du^2+dx^i dx^i}{u^2}
\ee
It is related to the Poincare patch by the coordinate transformation
\be
\tau = \arcsin \left( \frac{t}{z} \right)  \,, \qquad u = \sqrt{z^2-t^2}
\ee
after which we obtain the usual Poincare metric
\be
ds^2= \frac{-dt^2+dz^2 +dx^i dx^i}{z^2}
\ee
From this, we can easily see that the WdW patch only covers the portion of the Poincare patch where $z>|t|$, which is the causal development of the $t=0$ slice. This is shown in Fig. \ref{fig:2}.

\subsection*{New York Transformation}
As explained in section \ref{Yorktrans}, it is now easy to find the variation that yields $\delta K \propto h$ and $\delta h=0$, it is simply a $\tau$ translation. Performing the diffeomorphish $\tau\to\tau+2\alpha$, we get the metric
\be
ds^2 = -d\tau^2 + \cos^2\tau(1-2\alpha\tan \tau) \  \frac{du^2+dx^i dx^i}{u^2} 
\ee
 to first order in $\alpha$. We can now perform the change of coordinates to map back to Poincare coordinates. The metric is given by
 \bea 
 ds^2&= \frac{-dt^2+dz^2 +dx^i dx^i}{z^2} -2\alpha \Big[\frac{t^3}{z^2(z^2-t^2)^{3/2}}dt^2  \\
 &- \frac{t^2}{z(z^2-t^2)^{3/2}}dtdz+\frac{t}{(z^2-t^2)^{3/2}}dz^2+\frac{t}{z^2(z^2-t^2)^{1/2}}dx^idx^i\Big]
 \eea
One can easily see that
\bea
\delta h_{ab}&=0 \\
\delta K_{ab}&= \alpha h_{ab} 
\eea

It is worth looking at the behaviour of this metric near the AdS boundary. In Euclidean, we have the following variation of the boundary metric
\be
\delta_{Y} \gamma_{\mu \nu}  = 2 i \alpha \ \text{sign} (t_E) \delta_{\mu\nu} \,,
\ee 
which is a singular Weyl transformation of the boundary metric. It is quite elegant how the bulk regularizes the singularity of the boundary, which is similar to how the bulk regularizes the conical singularity in the replica trick \cite{Camps:2016gfs}. Note that there is a breakdown of the Fefferman-Graham coordinates in such scenarios. It is also easy to see how the Weyl transformation is generated using the HKLL procedure \cite{Hamilton:2006az}. 

\section{Symplectic form and volume for the large $h$ scalar condensate}
\label{app:scalarcond}

First we want to reproduce the bulk symplectic flux \eqref{eq:scalarvolume} via a boundary calculation. Since we are in a hyperbolic frame, the boundary two point functions to be used are not obvious, but we can read them off from the bulk to boundary propagator \eqref{eq:wdwbulktobndy} in the $\tau_E \rightarrow \pm \infty$ limit. The result is that\footnote{See \cite{Maldacena:2004rf} for a related discussion.} 
\beq
\langle O(u,x) O(u',x') \rangle =d_h \left( \frac{4 u u'}{(x-x')^2+(u+k u')^2} \right)^{2h} 
\eeq
Here, $k=-1$ if the two insertions are on the same side of the boundary manifold and $k=1$ if they are on different sides. 
The $k=-1$ case is easily understood from the boundary point of view as the flat space two point function with the $O(u,x) \mapsto u^{2h}O(u,x)$ transformation of operators under the Weyl rescaling $du^2+dx^2 \mapsto \frac{1}{u^2}(du^2+dx^2)$ of the metric. The $k=1$ case involves the same Weyl transformation along with a time reflection for one of the operators. The coefficient $d_h$ is not fixed by any of the above arguments; it is fixed by the particular normalization for sources that we use and must be determined by evaluating the bulk scalar on shell action with the bulk to boundary propagator \eqref{eq:wdwbulktobndy}. Now we want to calculate the boundary symplectic form \eqref{eq:qftkahler}, which in the present case reads as
\beq
i d_h\int \frac{du}{u^2} dx \int \frac{du'}{u'^2} dx' \left( \frac{4 u u'}{(x-x')^2+(u+u')^2} \right)^{2h} [\delta \lambda^-(u) \delta' \lambda^+(u')-\delta' \lambda^-(u) \delta \lambda^+(u')].
\eeq
Note that there is no need in this frame to restrict the regions of integration, the fact that the sources are on different hemispheres is just encoded in using the 2 point function with $k=1$.
We can easily do one of the $x$ integrals by the Schwinger parameter trick, and we can do the $u'$ integral again by saddle point for large $h$. The result is
\beq
4 d_h i \frac{\pi}{2h}\int dx \int \frac{du}{u^2} [\delta \lambda^-(u) \delta' \lambda^+(u)-\delta' \lambda^-(u) \delta \lambda^+(u)].
\eeq
Using \eqref{eq:boundaryscalarweyl} for the $\delta'$ transformed sources, the result agrees with \eqref{eq:scalarvolume}, provided $d_h=2\pi c_h^2 = h^2/(2\pi)$ to leading order in large $h$. 

We can explicitly verify that the backreacted volume agrees with \eqref{eq:scalarvolume}. We want to solve Einstein's equation with a stress tensor sourced by the profile \eqref{eq:largehsol}. To leading order in $1/h$, the solution can be written as
\bea
\label{eq:largehmetricansatz}
ds^2&=ds_{AdS}^2 \\
&+c_h^2 \cosh^2 \tau_E \Big[ \big( \frac{8\pi^2}{u^2}\lambda^-(u)\lambda^+(u)+b(u) \big) du^2-b(u) dx^2\Big] \\
&-\frac{ c_h^2 \pi^2}{2 h^2 u^2}\left( e^{(2-4h)\tau_E} \lambda^-(u)^2+ e^{(4h-2)\tau_E} \lambda^+(u)^2\right)[dz^2+dx^2],
\eea
where $ds_{AdS}^2$ is the Euclidean version of the WdW metric \eqref{WdWmetric}, and
\bea
b(u) &=\int^u_0 du' \frac{\int^{u'}_{0}du'' \frac{8\pi^2(\lambda^-(u'')\lambda^+(u'') )'}{u''}}{u'^2}. 
\eea
This gives the desired change in the volume. Note that the solution can be changed by a diffeomorphism $u\mapsto u+G(u)$, but this changes the volume only by a boundary term.

\section{Time dependence of the information metric in the $d=2$ thermofield double}
\label{app:infmetricTFD}

Let us pick a codimension one bulk surface given by the function $\rho(t_E)$ in the Euclidean version of the metric \eqref{eq:2dtfdmetric} (we imagine translational symmetry along $x$) and calculate the canonical momenta corresponding to the solution \eqref{mode}. The normal vector to this surface is
\beq
\partial_n=\frac{1}{\sqrt{\rho'(t_E)^2+\sinh^2 \rho}}\Big( \sinh\rho \partial_\rho-\frac{\rho'(t_E)}{\sinh \rho} \partial_{t_E}\Big),
\eeq
and we have
that the bulk formula \eqref{eq:infmetricbulk} for the information metric gives
\bea
\label{eq:marginalflux}
\int_\Sigma \sqrt{|h|}\partial_n \delta_h \phi &= 
c V_{bndy} \int_{-\frac{\pi}{2}-{\cal T}_E}^{\frac{\pi}{2}+{\cal T}_E} dt_E \int_{-\frac{\pi}{2}-{\cal T}_E}^{\frac{\pi}{2}+{\cal T}_E} dt_E' \int_{-\infty}^\infty dx \\
&\frac{2\cosh \rho(\rho'(t_E)\sin(t_E-t_E')+\cos(t_E-t_E')\cosh\rho \sinh\rho-\cosh x \sinh^2\rho)}{\big[ \sinh\rho \cos(t_E-t_E')-\cosh\rho \cosh(x)\big]^3}.
\eea
Here $V_{bndy}=\int dx$ is the boundary volume. The result should be independent of $\rho(t_E)$ as long as it is boundary anchored, because of the conservation of the symplectic flux. Let us confirm this explicitly, just for the sake of illustration. We take a functional derivative of \eqref{eq:marginalflux} giving
\bea
\label{eq:marginalflux2}
\delta \int_\Sigma \sqrt{|h|}\partial_n \delta_h \phi &= 
A V_{bndy} \int_{-\frac{\pi}{2}-{\cal T}_E}^{\frac{\pi}{2}+{\cal T}_E} dt_E \int_{-\frac{\pi}{2}-{\cal T}_E}^{\frac{\pi}{2}+{\cal T}_E} dt_E' \int_{-\infty}^\infty dx \left(\frac{\partial L}{\partial \rho}-\frac{d}{dt_E}\frac{\partial L}{\partial \rho'}  \right) \delta \rho \\& + A V_{bndy} \left[ \int_{-\frac{\pi}{2}-{\cal T}_E}^{\frac{\pi}{2}+{\cal T}_E} dt_E' \int_{-\infty}^\infty dx \frac{\partial L}{\partial \rho'} \delta \rho \right]_{-\frac{\pi}{2}-{\cal T}_E}^{\frac{\pi}{2}+{\cal T}_E}\\
L(\rho(t_E),\rho'(t_E))\equiv &\frac{2\cosh \rho(\rho'(t_E)\sin(t_E-t_E')+\cos(t_E-t_E')\cosh\rho \sinh\rho-\cosh x \sinh^2\rho)}{\big[ \sinh\rho \cos(t_E-t_E')-\cosh\rho \cosh(x)\big]^3}.
\eea
One can easily check that
\bea
\frac{\partial L}{\partial \rho}-\frac{d}{dt}\frac{\partial L}{\partial \rho'} &= -\frac{2\cos(t_E-t_E')\cosh x \sinh^2 \rho+(\cosh 2x-2)\sinh 2\rho}{(\cosh \rho \cosh x-\cos(t_E-t_E')\sinh \rho)^4} \\
&=\frac{d}{dx} \left(\frac{-2\sinh \rho \sinh x}{(\cosh \rho \cosh x - \cos(t_E-t_E')\sinh \rho)^3} \right),\\
&\Downarrow \\
\int_{-\infty}^{\infty} dx &\left( \frac{\partial L}{\partial \rho}-\frac{d}{dt}\frac{\partial L}{\partial \rho'} \right)=0.
\eea
Therefore, \eqref{eq:marginalflux} is a boundary term for any $\rho(t_E)$ as it should (up to a $\rho(t_E)$ independent functional).
In particular, since we only care about surfaces where at the boundaries of the surface $\rho \rightarrow \infty$, we may take directly $\rho(t_E)\rightarrow \infty$ in  \eqref{eq:marginalflux} and recover
\beq
\int_\Sigma \sqrt{|h|}\partial_n \delta \phi =A V_{bndy} \int_{-\frac{\pi}{2}-{\cal T}_E}^{\frac{\pi}{2}+{\cal T}_E} dt_E \int_{-\frac{\pi}{2}-{\cal T}_E}^{\frac{\pi}{2}+{\cal T}_E} dt_E' \int_{-\infty}^\infty dx \frac{-2}{(\cos(t_E-t_E')-\cosh x)^2}.
\eeq
We notice that the integrand is the marginal correlator, and this formula of course agrees with the one considered in \cite{MIyaji:2015mia} from the boundary point of view.

\bibliographystyle{utphys}
\bibliography{volume}

\end{document}